\documentclass{emulateapj}
\usepackage{natbib,graphicx,url}
\usepackage{color}
\shorttitle{Lensing Effects from Halo-to-Halo Variation}
\shortauthors{Chen, Koushiappas \& Zentner}

\def\fsub{f_{\mathrm{sub}}}
\def\Vmax{V_{\mathrm{max}}}
\def\Mh{M_{\mathrm{h}}}
\begin{document}

\title{The Effects of Halo-to-Halo Variation on Substructure Lensing}
\author{Jacqueline Chen\altaffilmark{1,3}, Savvas M. Koushiappas\altaffilmark{1,4}, and Andrew R. Zentner\altaffilmark{2,5}}
\altaffiltext{1}{Department of Physics, Brown University, 182 Hope St., Box 1843, Providence, RI 02912} 
\altaffiltext{2}{Department of Physics and Astronomy, University of Pittsburgh, Pittsburgh, PA 15260}
\altaffiltext{3}{{\tt Jacqueline\_Chen@brown.edu}} 
\altaffiltext{4}{{\tt koushiappas@brown.edu}} 
\altaffiltext{5}{{\tt zentner@pitt.edu}}

\begin{abstract}
We explore the halo-to-halo variation of dark matter substructure in galaxy-sized dark matter halos, focusing on its implications for strongly gravitational lensed systems.  We find that the median value for projected substructure mass fractions within projected radii of 3\% of the host halo virial radius is approximately $f_{\rm sub} \approx 0.25\%$, but that the variance is large with a 95-percentile range 
of $0 \le f_{\rm sub} \le 1\%$.  We quantify possible effects of substructure on quadruply-imaged lens systems using 
the cusp relation and the simple statistic, $R_{\rm cusp}$.  We estimate that the probability of obtaining the large values of the $R_{\rm cusp}$ which have been observed from substructure effects is roughly $\sim 10^{-3}$ to $\sim 10^{-2}$.  We consider a variety of possible correlations between host halo properties and substructure properties in order to probe possible sample biases.  In particular, low-concentration host dark matter halos have more large substructures and give rise to large values of $R_{\rm cusp}$ more often.  However, there is no known observational bias that would drive observed quadruply-imaged quasars to be produced by low-concentration lens halos.  Finally, we show that the substructure mass fraction is a relatively reliable predictor of the value of $R_{\rm cusp}$.  
\end{abstract}


\section{Introduction}

In the standard Cold Dark Matter ($\Lambda$CDM) cosmology, structure forms hierarchically.  Small dark matter halos form first and merge to form larger objects.  The process of merging and dynamical evolution leaves remnants of smaller halos (called subhalos) within larger host halos.  We refer to the population of subhalos as {\it substructure}.  The presence of substructure in galaxy-sized dark matter halos is an important test of cold dark matter.  While the distribution and number of galaxies and clusters of galaxies is in agreement with the generic predictions of CDM, at sub-galactic scales numerical simulations predict an abundance of dark matter substructure which has not been detected by the presence of baryonic matter (stars \& gas), a situation known as the ``missing satellites problem" \citep{klypin_etal99,moore_etal99,diemand_etal07,springel_etal08}.  Detecting dark matter subhalos and putting constraints on the substructure mass function is a key step toward the identification of the dark matter particle \citep[see, e.g.,][]{moustakas_etal09}. 

One of the most promising avenues to measuring substructure in galaxies is via strong gravitational lensing.  As light from a distant object passes an intervening galaxy, it may be bent sufficiently for multiple images of the source to be observed.  Strong lens systems with quasar sources and galaxy-sized lenses have been shown to be sensitive to satellite galaxy-sized dark matter subhalos in the lensing halo\footnote{Since gravitational lensing is sensitive only to the matter distribution in the lens, the term ``lens halo" or ``lensing halo" is used to denote the combination of the dark matter halo and the baryonic components within that dark matter halo.}\citep{mao_schneider98,metcalf_madau01,metcalf_zhao02,dalal_kochanek02,chiba02,kochanek_dalal04}.  Substructure in the lensing halo may be observed as perturbations to the image magnifications (flux ratio anomalies), image positions \citep[e.g.,][]{metcalf_madau01,chen_etal07}, and/or time delays in systems with time variable sources \citep{keeton_moustakas09}.  

The most well-explored approach to substructure lensing is the study of anomalous flux ratios in multiply-imaged systems.   
Anomalous flux ratios are sensitive to the projected mass fraction in substructure, $\fsub$, at projected radii where lensed images are present, $\sim$3\% of the projected virial radius of the lens halo.  Results, however, have not been definitive.   \citet{dalal_kochanek02} find that $f_{\rm sub}=2\%$ with $0.6\% < f_{\rm sub} < 7\%$ within 90\% confidence intervals.  This is marginally consistent with the results of numerical simulations.  \citet{metcalf_amara10} also find a result which is generally consistent with numerical simulations.   

Several studies have shown that $f_{\rm sub}$ is small ($\lesssim 1\%$) but with large variance.  For example, \citet{zentner_bullock03} used semi-analytic simulations of a large number of halos to argue that substructure 
mass fractions at projected radii of 3\% of halo virial radii should be $f_{\rm sub} \approx 0.6\%$ on average and 
$f_{\rm sub} \lesssim 1\%$ for $95\%$ of halos.  \citet{mao_etal04} found $f_{\rm sub} \lesssim 0.5\%$ using 12 halos at galaxy, group and cluster masses.  \citet{xu_etal09} found $.01\% \leq f_{\rm sub} \leq .7\%$, using 6 halos from the Aquarius simulation.  Using 26 simulated halos, \citet{zentner06} showed that $\fsub < 1\%$ along most lines of sight, but that there is a large variance due, in part, to the orientation of the halo with respect 
to the line of sight.  

Other studies have suggested that lensing observations are inconsistent with expectations from numerical simulations.  Using simple models for substructure, \citet{chen09} found that observations support the presence of more substructure than expected from simulations.  \citet{maccio_etal06} and \citet{amara_etal06} each compared observations to a single simulated halo and found more and greater magnification perturbations than can be accounted for in their simulated halos.  In addition, the results of \citet{dalal_kochanek02} and \citet{metcalf_amara10} (which are consistent with simulations) use substructure models that make it  possible that their results underestimate the value of $f_{\rm sub}$.  

Future, more definitive, results will require better characterizations of sources of error, more observational data, and better predictions for CDM substructure. In this work we focus on this last element. More specifically, as the initial density field is unknown, predictions for substructure populations in numerical simulations are specific to a particular realization of the simulated halo. Halo-to-halo variation, which should be present, is essential in determining the expected variance about each measurement.  Quantifying this variance can allow more accurate error assessments and potentially help bring observational results into better agreement with theoretical predictions. In addition, understanding the variance of the substructure population may provide valuable insight into previously unexplored biases in the lensing samples.  

In this work, we characterize the halo-to-halo variation of substructure populations in galaxy-sized dark matter halos and investigate how substructure properties correlate with other properties of dark matter halos.  We study the effects of halo-to-halo variation to mock lensing observations, focusing on a simple statistic that quantifies violations of the cusp relation. We employ a semi-analytic method to generate a large number of substructure populations. We describe this method in Sec.~\ref{sec:sa} and describe basic properties of halo substructure in 
Sec.~\ref{sec:substat}.  We use our substructure models to generate strong lensing simulations in Sec.~\ref{sec:strong}. We analyze our simulations and compare the results to observations in Sec.~\ref{sec:compare}, we discuss caveats in Sec.~\ref{sec:caveats}, and we conclude in Sec.~\ref{sec:conclusions}. Throughout this paper, we work in a flat cosmological model with $\Omega_{dm} = 0.228$, $\Omega_b = 0.0227$, $h=0.71$, and $\sigma_8 = 0.81$ as favored by the five-year Wilkinson Microwave Anisotropy Probe results \citep{2009ApJS..180..330K}.

\section{Semi-Analytic Model}
\label{sec:sa}

In order to generate a large statistical sample of host halos, we use a semi-analytic technique \citep{zentner_etal05}. This technique allows us to approximate the hierarchical assembly of a dark matter halo analytically, and thus avoid the prohibitive computational cost associated with numerous, high resolution numerical simulations \citep[see also][]{taylor_babul01,taylor_babul04}.

The semi-analytic approach employs a variation of the extended Press-Schechter formalism to generate merger histories of host dark matter halos \citep{1991ApJ...379..440B,1993MNRAS.262..627L,1999MNRAS.305....1S,2007IJMPD..16..763Z} and an approximate analytic approach to the dynamical evolution of subhalos in an evolving host halo.  The details of this method are given in \cite{zentner_etal05}, including a number of comparisons to self-consistent numerical simulations. 
This technique has been tested for subhalos of mass  $M > 10^{-4} M_{\rm h}$, where $M_{\rm h}$ is the mass of the host halo, 
and produces subhalo abundances and spatial distributions in good agreement with $N$-body simulations.  
Though this approach is approximate, it allows us to compute a large number of realizations of a host halo at a 
subhalo resolution which is inaccessible at present with state-of-the-art numerical simulations. 
In addition, this procedure can be used to produce physically motivated extrapolations to numerical simulations 
\cite[see][]{2006ApJ...639....7K,Koushiappas:2003bn,2010PhRvD..82h3504K}.

We generate 200 realizations of a host halo of mass $M_{\rm h} = 1.26 \times 10^{12} h^{-1} M_\odot$. We track the accretion and dynamical evolution of subhalos of mass greater than $10^{-5} M_{\rm h}$. The mean properties of substructure in the sample of realizations is in agreement with numerical simulations \citep[see Fig. 1 in][]{2010PhRvD..82h3504K}. The average number of subhalos with $\Vmax > 4$ km/s within the inner 200 kpc of the halo is 2481 with a 68 percentile range of $[1964-3007]$, which is consistent with the 2469 subhalos found within the same radius of a numerical simulation of a similar-size halo \citep{2008ApJ...686..262K}.  In addition, the scatter in the number of subhalos from halo-to-halo variation is consistent with a Poisson scatter added in quadrature to an intrinsic scatter of about 20\%  \citep[as derived from recent numerical simulations;][]{boylan-kolchin_etal10}.

Each realization represents a possible subhalo population within the host halo. The distribution, structure and properties of the subhalo population differ between realizations due to the statistical nature of the initial density field and the complexity of any individual subhalo's orbit. The output of each realization contains the concentration of the host and a list of all subhalos, including their evolved structural parameters -- total bound mass, scale radius, and tidal radius -- and their locations within the host \citep[see][]{zentner_etal05}. The individual merger history of each host influences the abundance and properties of its substructure population.

\section{Substructure Statistics}
\label{sec:substat}

\begin{figure}[h]
\centering
\resizebox{3.2in}{!}	{\includegraphics{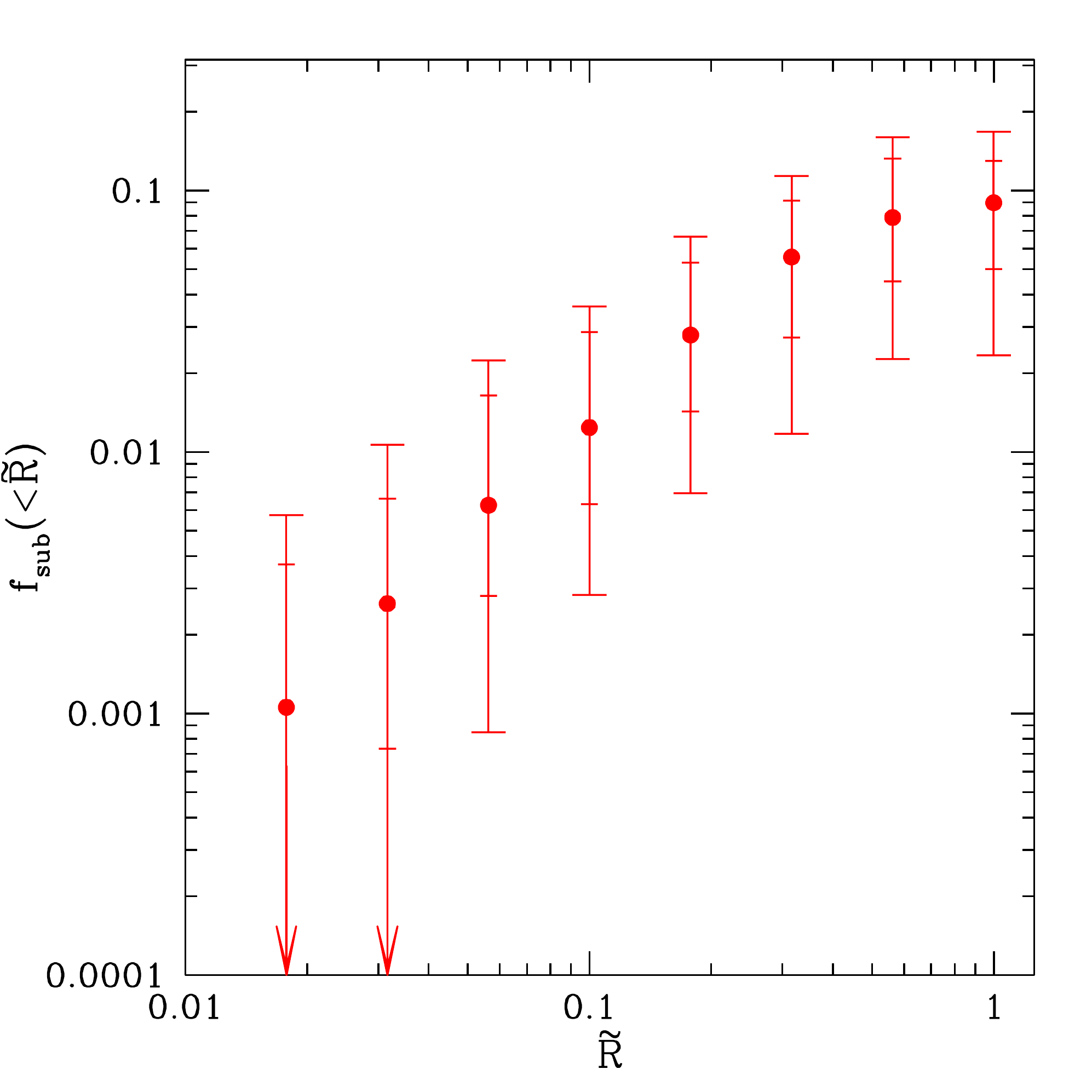}}\\
\resizebox{3.2in}{!}	{\includegraphics{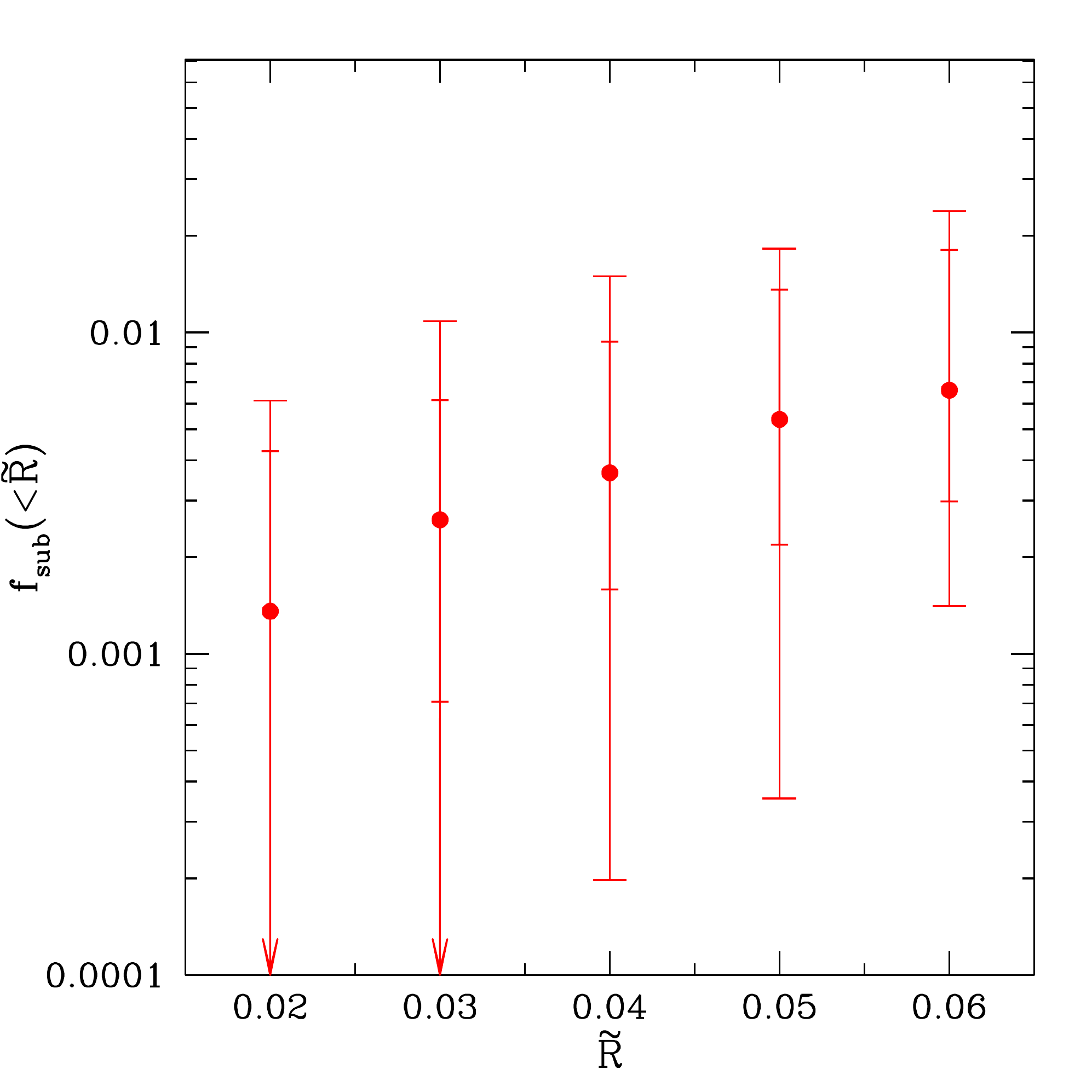}}
\caption{The projected substructure mass fraction as a function of radius ($\tilde{R} = R/R_{\rm h}$) for subhalos with mass greater than $10^{-5}$ times the mass of the host halo.  The median in the sample of 200 realizations is shown as a point, the 68 percentile is shown with narrow errorbars, while the 95 percentile is shown with wide errorbars.  The large spread of $\fsub$ at large radii is an outcome of the distribution of merger histories of the host halos. The top panel shows the projected substructure mass fraction at radii $\tilde{R} = [0-1]$, while the bottom panel shows the projected substructure mass fraction in the inner regions of the halo, which is relevant to lensing studies.  
\label{fig:Fig1}}
\end{figure}

We construct a two-dimensional projection of the subhalo distribution by projecting the three-dimensional density along an 
arbitrary direction.  The three-dimensional density profile of each subhalo is described by a Navarro-Frenk-White profile \citep[NFW;][]{navarro_etal97}, $\rho(r) = \rho_s / [x (1+x)^2]$, where $x=r/r_s$, $r_s$ is the scale radius, and $r$ is the three-dimensional distance from the halo center.   Each subhalo has a projected distance from the center of the halo, which we call $R$. 
For simplicity we define a dimensionless projected distance for each subhalo as $\tilde{R} = R / R_{\rm h}$, where $R_{\rm h}$ is the virial radius of the host halo. This two-dimensional projection of the subhalo population is used in the mock lensing simulations we describe below. 
We also utilize a dimensionless subhalo mass $\tilde{M}$ defined such that the mass of a subhalo is $M= \tilde{M} M_{\rm h}$, 
where $M_{\rm h}$ is the mass of the host halo.

We can glean information about the variation in the distribution of substructure by calculating the subhalo mass fraction within a projected radius. The projected mass fraction within the inner $\tilde{R}$ fraction  of the virial radius of the host in a given subhalo mass bin is
\begin{equation} 
\fsub(<\tilde{R}) = \frac{1}{\tilde{M}_{\rm h}(<\tilde{R})} \sum_i \tilde{M}_i(\tilde{R}_i < \tilde{R}) \label{eqn:fsub}
\end{equation} 
where 
\begin{equation} 
\tilde{M}_h(<\tilde{R}) = \frac{c^2}{\mathcal{F}(c)} \int_0^{\tilde{R}} \tilde{R}'\int_0^{{\ell_{\mathrm{max}}}} \frac{1}{\tilde{r} [ 1 + \tilde{r} ]^2} \, \,d \tilde{\ell} \,\, d\tilde{R}'
\label{eq:mr}
\end{equation} 
is the projected mass within the inner $\tilde{R}$ fraction of the virial mass of the host halo. In Eq.~\ref{eqn:fsub}, the sum over subhalo masses is over all subhalos $i$ whose projected radial position $\tilde{R}_i$ is less than the projected host position $\tilde{R}$, independent of the projected size of the subhalo. 
In Eq.~\ref{eq:mr}, $\tilde{r} = c  r / R_{\rm h} = \sqrt{ \tilde{R}^{\prime 2} + \tilde{\ell}^2}$, the upper integral in the line of sight element is $\tilde{\ell}_{\mathrm{max}} = \sqrt{1 - \tilde{R}^{\prime^2}}$, and $\mathcal{F}(c) = \ln( 1 + c ) - c / ( 1 + c)$, where $c$ is the concentration of the host halo.  Note that values of $\fsub \geq 1$ are possible.  This may be the result of two effects:   $\tilde{M}_{\mathrm{h}} (< \tilde{R})$ from Eq. \ref{eqn:fsub} only refers to the smooth mass of the host halo and not the total mass, and we use the full subhalo mass even when most of the subhalo's bound mass falls outside of the relevant region.

\begin{figure}
\centering
\resizebox{3.4in}{!}	{\includegraphics{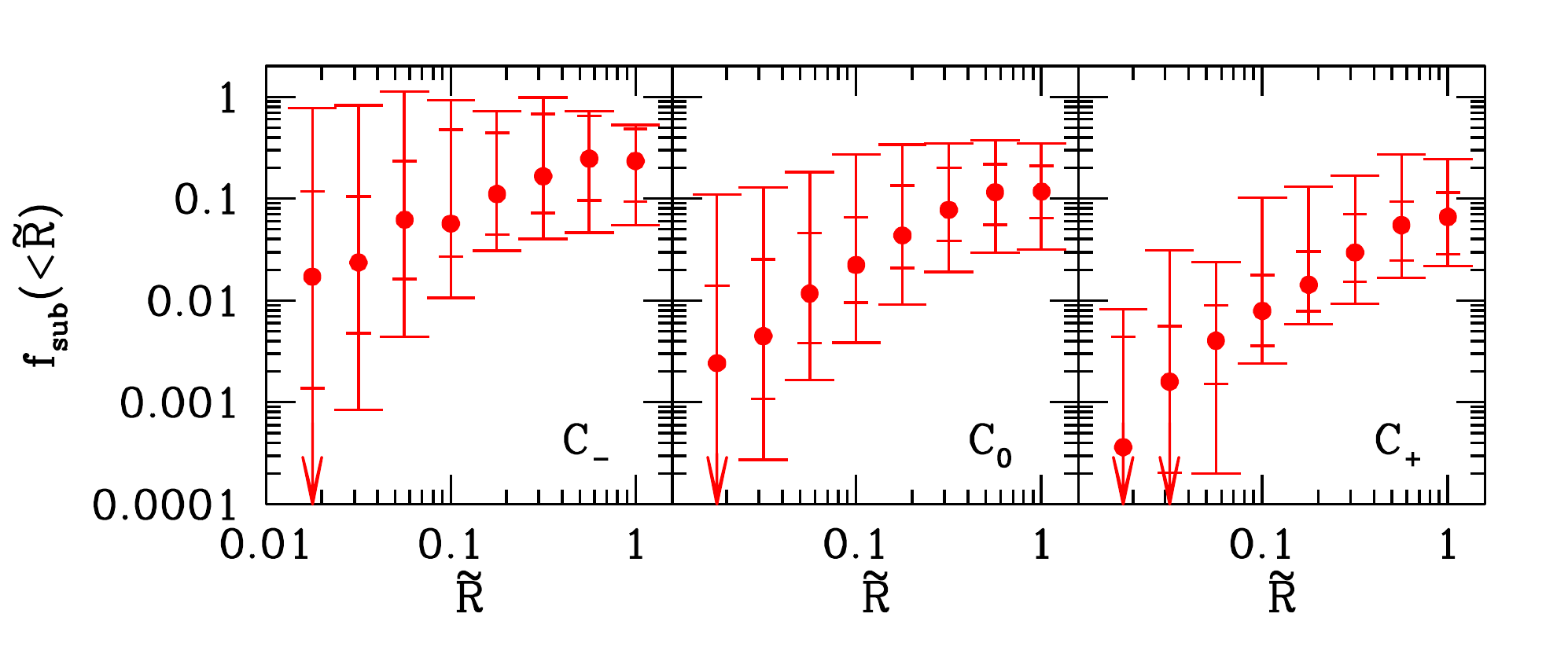}}
\resizebox{3.4in}{!}	{\includegraphics{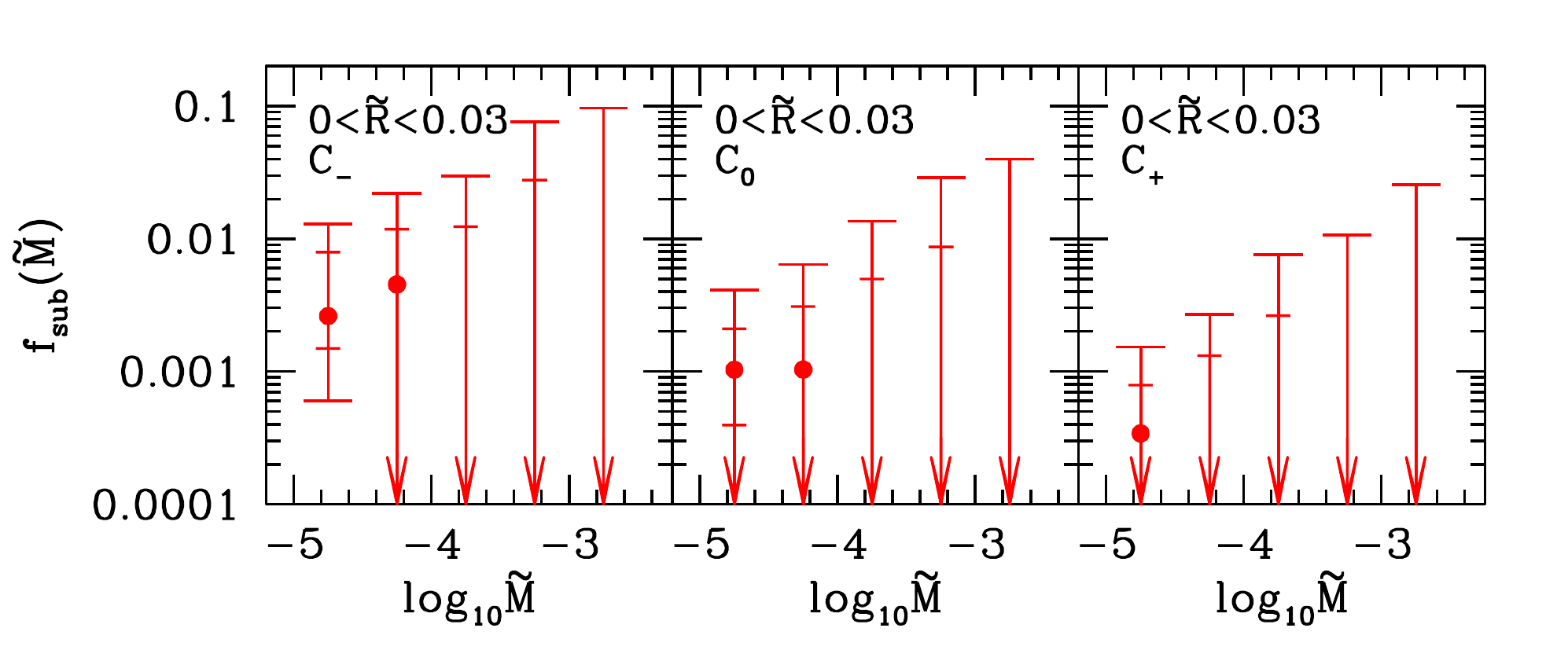}}
\caption{{\em Top}: The projected substructure mass fraction as a function of radius ($\tilde{R} = R/R_{\rm h}$) for subhalos with mass greater than $10^{-5}$ times the mass of the host halo. Host halos with concentrations in the 68 percentile of the mean have substructure mass fractions which are of order $\sim 10\%$, while lower and higher concentration hosts have a substructure mass fraction which is higher and lower, respectively. See text for details. {\em Bottom}:  The mass fraction within the projected inner 3\% radius of objects in a given mass bin for hosts with low (left panel), mean (middle panel) and high (right panel) concentration. In both figures, the y-axis errorbars represent the same percentile range as in Fig.~\ref{fig:Fig1}.  Points depict the median of the distribution. Note that as the concentration of the host increases, the fraction of mass contributing to the inner 3\% of the projected radius becomes negligible. As a result, only hosts with low concentration are likely to have rich substructure in the projected inner 3\% of their radius. 
\label{fig:Fig2}}
\end{figure}

In the top panel of Fig. \ref{fig:Fig1} we show the projected cumulative substructure mass fraction $\fsub$, in the complete sample of 200 realizations as a function of radius. The median value of $\fsub$ at $\tilde{R}=1$ is 9\% for subhalos with mass greater than $10^{-5} \Mh$, with a 95 percentile in the range $[2-17]\%$. Note that the spread about the median increases as $\tilde{R}$ increases. This is due to the fact that the outer regions of halos are dominated by recently accreted subhalos, while the subhalo population of the inner regions is old and highly evolved.  The recent accretion of subhalos is an outcome of the individual accretion history of each host halo, and the large spread about the median in the outer regions mirrors the range of recent accretion histories.

Relevant to lensing studies is the projected substructure mass fraction within the inner 3\% of the virial radius of the host halo. The bottom panel in Fig.~\ref{fig:Fig1} shows $\fsub$ in the inner regions. At $\tilde{R} = 0.03$, the median value of $\fsub$ is 0.25\% with a 95 percentile in the range $[0-1]\%$, meaning that only 2.5\% of the realizations have a projected substructure mass fraction greater than 1\% in the inner 3\% of the virial mass of the host halos.

In order to explore the relationship between the host halo properties and the substructure lensing effects, we distribute realizations into three groups. In the first group, which we call $C_0$, we include host halos with a concentration $c$ (c.f., Eq. \ref{eq:mr}) within the 68\% range of the mean concentration of the complete sample, $\langle c \rangle = 9.7$. This includes host halos with concentrations in the range $ 6.7 \le c \le 13.4$. We then collect the upper 16\% of the concentration range ($c> 13.4$) in group $C_+$, and the lower 16\% range ($c<6.7$) in group $C_-$.  As they form earlier, subhalos present in high concentration hosts are expected to be fewer and with more concentrated subhalo density profiles relative to subhalos present in low concentration hosts.  The inverse relation between substructure abundance and halo concentration has also been seen in the simulations of \citet{ishiyama_etal08} and \citet{ishiyama_etal09}, where they measure the abundance of subhalos around dark matter halos of $\sim10^{12} M_{\sun}$ and larger.  Grouping the hosts (and their substructure populations) allows us to investigate the effects of host concentration on the expected lensing signal from its substructure. 

The cumulative substructure mass fraction for halos with mass greater than $10^{-5}$ the mass of the host is also dependent on the concentration of the host halo. In the top panel of Fig.~\ref{fig:Fig2} we show the median and 68 \& 95 percentiles of the distribution of mass fractions found in our sample. Host halos with concentration in the 68 percentile about the median concentration of all realizations (middle figure) have a substructure mass fraction which is of order $\sim 10\%$, and a 68 percentile of $\sim[5-20]\%$ (the 95 percentile range is $\sim[3-35]\%$). Higher concentration hosts have a smaller fraction (of order $\sim 6\%$ and a 68 percentile range of $\sim[3-10]\%$, and a 95 percentile range of $\sim[2-25]\%$), while low concentration hosts have a higher substructure mass fraction (median of $\sim 22\%$, and a 68 percentile range of $[9-50]\%$ and a 95 percentile of $\sim[4-55]\%$).  
Systems of low concentration (and subsequently large values of $\fsub$) are systems which typically contain recent large mergers.  Such systems may appear as interacting galaxy pairs or groups in observations.  

In the bottom panel of Fig.~\ref{fig:Fig2} we show the projected mass fraction within the inner 3\% of the projected virial radius (most relevant to lensing studies), for a number of mass bins.\footnote{This particular choice of mass bins is for illustrative purposes only.} We find that the abundance of subhalos with masses greater than $\sim$ few times $10^{-4}$ the mass of the host is negligibly small:    even though the upper 95 percentile of the mass fraction increases with mass in the panels, the median and the upper 68 percentile drop to zero (as does the number abundance) as mass increases.  Therefore, large subhalos contribute little to projected mass fractions on average, although, in the rare cases in which a large subhalo is present, the projected mass fraction is correspondingly large.  This deficit of large subhalos is independent of the host halo concentration. However, on smaller scales (masses of order $\tilde{M} \sim 10^{-4}-10^{-5}$), the effects of the host concentration are prominent.  For low-concentration host halos, the median mass fraction is $f_{\rm sub} \approx 0.3\%$ with an upper 68 percentile of $f_{\rm sub} \approx 0.8\%$ 
(95 percentile is $f_{\rm sub} \approx 1.3\%$), within the projected inner 3\% of the virial radius.  
In contrast, the high-concentration host halos have a median of $\fsub \approx 0.035\%$, and with an upper 95 percentile of less than 0.2\%. 

The origin of these differences between low- and high-concentration halos is the same as alluded to earlier.  High-concentration halos generally form earlier than low-concentration counterparts, 
undergoing the majority of their mergers at higher redshift.  
As a consequence, the merging subhalos evolve for a relatively longer period within the dense environments of their hosts in high-concentration halos.  This 
allows more time for orbital decay and mass loss.  The increased propensity to 
lose mass in high-concentration halo is compounded by the fact that tidal forces 
are moderately stronger in high-concentration host halos at fixed, total bound 
mass.  These effects, working together, render the survival rate of mass bound 
to halo substructure within high-concentration hosts smaller than in low-concentration 
hosts.  The opposite is true of substructure in low-concentration host halos, so low-concentration 
host halos tend to have higher substructure mass fractions.  These concentration-dependent effects are 
consistent with those found and discussed in \citet{zentner_etal05} and \citet{wechsler_etal06}.

\begin{figure}[h]
\centering
\resizebox{3.6in}{!}	{\includegraphics{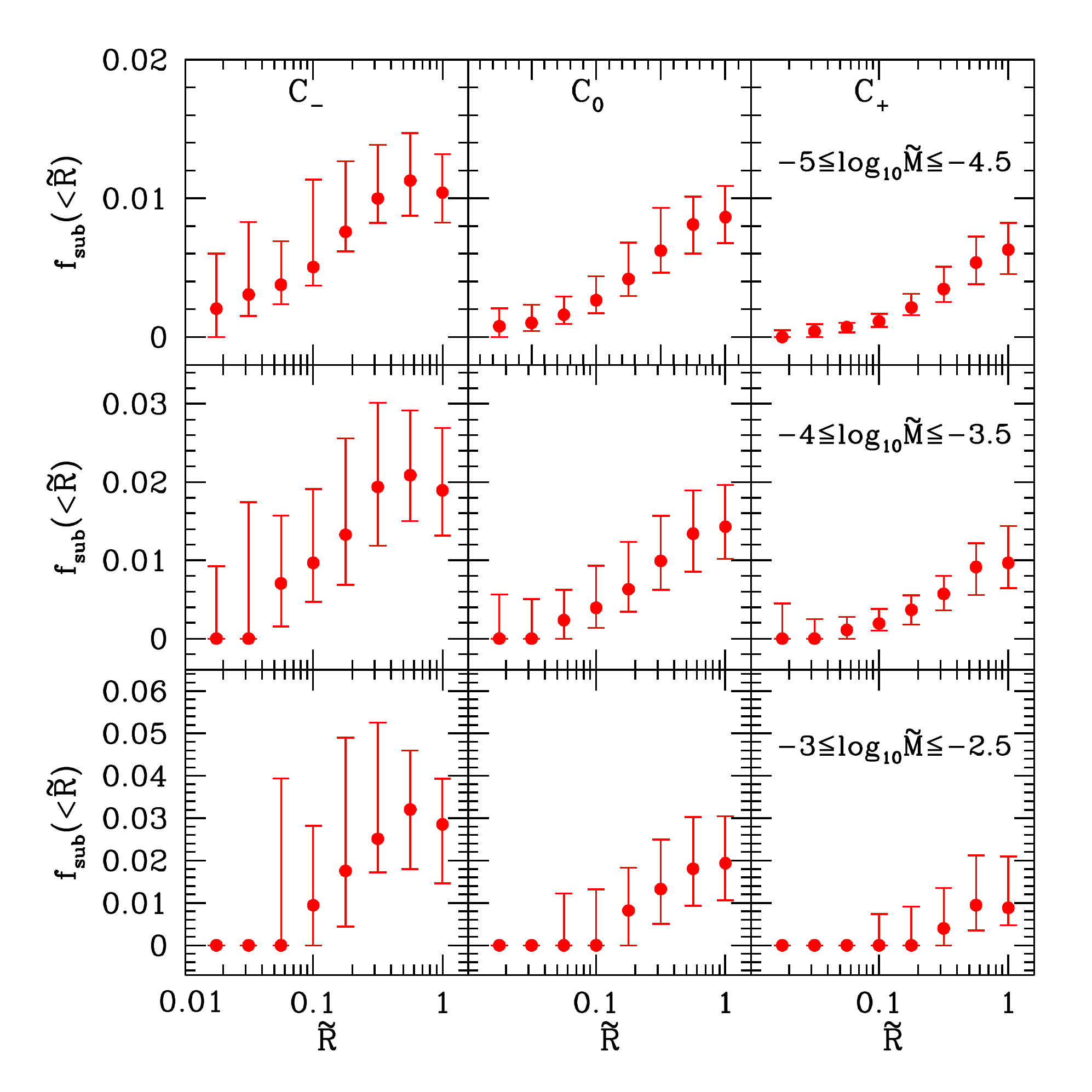}}
\caption{The mass fraction in substructure as a function of projected radius ($\tilde{R} = R/R_{\rm h}$) for 3 representative mass bins  (rows) for low, mean, and high concentration hosts (columns). Each row corresponds to the shown mass range of subhalos as a fraction of the host halo shown on the far right panel. The y-errorbars correspond to the 68 percentile range, and the median of the distribution is shown in points. See text for details.  
\label{fig:Fig3}}
\end{figure}

In Fig.~\ref{fig:Fig3} we summarize these effects by showing the projected substructure mass fraction as a function of radius for the three concentration groups of host halos and three representative subhalo mass bins.  It is apparent that host halos with low concentrations have a higher overall projected substructure mass fraction than halos with high concentration (see also top panel of Fig.~\ref{fig:Fig2}).  In addition, massive subhalos (subhalos with mass greater than $10^{-4}$ the mass 
of the host) tend to a median substructure mass fraction close to zero. However, the upper 68 percentile range is increasing relative to the 68 percentile of lower mass subhalos. For example, for masses $\sim 10^{-4}$ of the mass of the host, the upper 68 percentile implies that up to 16 percent of the low concentration halos have substructure mass fractions which are {\it greater} than 3 percent (with 2\% 
of the low concentration halos having a fraction greater than $\sim 7.5\%$). The corresponding fractions for subhalos with mass $\sim 10^{-5}$ the mass of the host are 1.2 and 3 percent (for the top 16 percentile and 2 percentile of hosts respectively). 

In addition, in high concentration hosts, the 68 percentile range increases with projected radius. The inner regions of these halos have evolved for a considerable amount of time, while the outer parts reflect more recent accretion.  On the other hand, in host halos with low concentrations, there has been, on average, less time for halo substructure to evolve.  Moreover, recent accretion of subhalos with either highly elongated orbits or with orbits that decay rapidly due to dynamical friction can deposit large amounts of substructure at either small or large halo-centric radii.  The consequence of this is that the range of substructure mass fractions extends to much higher values at all mass bins in low concentration hosts.

\section{Strong Lensing Simulations}
\label{sec:strong}

We create mock lens systems with characteristics typical of observed lens systems, and we include the substructure populations derived in the semi-analytic catalogs (see previous section).  We choose a lensing galaxy at $z=0.5$ and a source quasar at $z=2$ and model the lensing halo as a singular isothermal ellipsoid (SIE) with projected surface density,
\begin{equation}
\kappa(\xi) = \frac{\Sigma(\xi)}{\Sigma_{\rm crit}} = \frac{b}{2 \xi}, \label{eq:sie}
\end{equation}
where $\xi$ is the elliptical coordinate, $\xi^2 =x^2+y^2/q^2$, and $\Sigma_{\rm crit}$ is the critical surface density for lensing.  In the case of a spherical lens, $q=1$ and $b$ is the Einstein radius of the model.  In addition to the smooth lens model, an external shear, $\gamma$, is applied to account for nearby structure, such as a group of galaxies.  $\theta_\gamma$ is the angle between the shear and the major axis of the smooth lens model.

The host halo in the semi-analytic simulations is based on a Milky Way-sized dark matter halo.  It differs from lensing halos in a few respects.  First of all, it has no baryons.  The density profile of the Milky Way in dark matter only simulations matches a NFW profile.  This profile is shallower than isothermal in the inner portions and steeper than isothermal in the outer portions.  Real galaxies, however, contain baryons, and observations have shown that lensing halos (baryons + dark matter) follow a roughly isothermal profile near the halo center, where strong lensing is sensitive to substructure  \citep{treu_etal06}.  In addition to the differences in density profile, we might expect that models with baryons would have less substructure than our semi-analytic simulations, as the presence of the host galaxy makes tidal forces stronger.   On the other hand, this effect may be small as subhalos that are projected close to the halo center may be located at considerably larger physical distances.  

The host halo for the substructure catalogs also differs with respect to typical lensing halos in a few, more subtle ways:  the total mass, the formation and merger history, and the epoch of observation.  The host halo is a Milky Way-sized object, while lensing halos are mostly those which contain large, elliptical galaxies.  We might expect that realistic large, lensing halos might have somewhat more substructure than Milky Way-sized galaxies, as larger halos form later \citep{gao_etal04,giocoli_etal10}.  \citet{gao_etal04}, \citet{zentner_etal05}, and 
\citet{bosch_etal05} suggest that the number of substructures scales with host mass as 
${\rm d}N/{\rm d}\ln \tilde{M} \propto M_{\rm h}^{0.08}$, which would correspond to only a 15\% increase in substructure compared to our simulations.  A similar effect is possible as we use substructure catalogs at $z=0$ instead of at $z\sim [0.5-1]$ where lensing galaxies are found, since larger fractions of substructure might be expected at earlier epochs \citep{madau_etal08}.  These effects, however, are likely to be significantly smaller than the effect of halo-to-halo variation, as shown by \citet{xu_etal09} and \citet{zentner_etal05}.  

Overall, the effect of baryons on halo structure remains an important question, and the optimal method of placing dark matter (DM) substructure in a lensing halo remains unclear.  It might be possible to construct a sophisticated model that would connect the properties of the host halo to those of the lens halo, but it is neither relevant nor necessary for a general study, as is this work.  Further, treating the host halo and the lensing halo rather independently is logical because strong lenses have significant surface density contributions from baryonic components of galaxies and because lensing halos are described in the data well by isothermal models with a narrow range of dispersion \citep{treu_etal06}.  In the absence of a model for these effects and all the effects discussed above and in order to be conservative in our calculations, we preserve (1) the projected distance from the center of the halo (i.e., $\tilde{R}$ from the previous section) and (2) the fractional mass of substructure from our semi-analytic model in our lensing calculations.  In addition, the host halo mass, $M_{\rm h}=1.8\times10^{12} M_{\sun}$, is scaled to that of a more typical lensing halo, $M_{\rm lens}=10^{13} M_{\sun}$.  Thus,  the mass of each subhalo is $M_{\rm lens, sub} = (M_{\rm lens}/M_{\rm h}) M_{\rm sub}$ and all scale lengths are $x_{\rm lens, sub} = (M_{\rm lens}/M_{\rm h})^{1/3} x_{\rm sub}$, where $x$ denotes either the tidal radius or the scale radius, and both $\tilde{R}_{\rm lens} = \tilde{R}_{\rm}$ and $f_{\rm lens,sub} = f_{\rm sub}$.   

Using this model, the Einstein radius of a singular isothermal sphere (SIS) lensing halo at $z=0.5$ can be estimated from Eq. \ref{eq:sie}, $b=0.586\arcsec$.  This is generally consistent with observed lenses which have Einstein radii $\sim 1 \arcsec$ \citep[e.g.][]{browne_etal03}.  It is also a reasonable value for any likely lensing halo.  The Einstein radius of any SIS can be calculated from $b = 4\pi (\sigma/c)^2 D_{\rm ls}/D_{\rm os}$, where $D_{\rm ls}$ is the angular diameter distance from the lens to the source, $D_{\rm os}$ is the angular diameter distance from the observer to the source, and $\sigma$ is the 1-d velocity dispersion.  We expect lensing halos to have $\sigma=[200-350]$ km/s \citep{treu_etal06}, which corresponds with  $b=[0.366-1.120]\arcsec$.  In Table \ref{tab:sim}, we show the model parameters for the lens halo and environment. We refer to these as the macromodel parameters. 
   
\begin{table}[h]
\begin{center}
\begin{tabular}{c|c|c|c|c}
\hline
Name & $b$ & $q$ & $\gamma$ & $\theta_{\gamma}$ (rad) \\ 
\hline
\hline
M1 &  $0.586\arcsec$ & 0.7 & 0.2 & $\pi/3$ \\
M2 &  $0.586\arcsec$ & 0.7 & 0.1 & $\pi/6$  \\
M3 &  $0.586\arcsec$ & 0.9 & 0.2 & $\pi/3$ \\
\hline
\end{tabular} 
\caption{Simulation Parameters}
\label{tab:sim}
\end{center}
\end{table}

We model subhalos as truncated NFW profiles.  There is no closed form to calculate the image position deflections from NFW profiles with a sharp cut-off at the truncation radius.  We use one of the modified, truncated forms provided in \citet{baltz_etal09}, where
\begin{equation}
\rho(r)=\frac{\rho_s}{x(1+x)^2}  \frac{\tau^2}{\tau^2+x^2},
\end{equation}
where $\tau=r_t/r_s$ and $r_t$ is the truncation radius.  This form falls off as fast as $r^{-5}$ beyond the truncation radius and the relevant lensing quantities can be calculated simply.

Using the combination of the three different macromodels and the set of 200 substructure realizations, we perform lensing simulations using an inverse ray-tracing code.  The position of the source, $\vec{\beta}$, and the images, $\vec{\theta}$, are related by the lens equation, 
\begin{equation} 
\vec{\beta} = \vec{\theta} - \nabla \phi(\vec{\theta}), 
\end{equation}
where $\phi$ is the lensing potential. In the case of circular symmetry in the lens, $\nabla \phi_R \propto M(<R)$, where $M(<R)$ is the mass of the lens enclosed by a cylinder of radius $R$.  The lens equation is multi-valued and several image positions may correspond to a single source position.  Thus, it is solved most simply by establishing a grid of image positions and calculating the source position for each point on the grid.  For our simulations, the resolution on the image plane is 0.469 mas.  

Every set of three points in the image plane defines a triangle and corresponds to a triangle in the source plane.  The magnification of an image can be calculated by the ratio of the area of the image plane triangle to the area of the source plane triangle.  A common assumption -- one also used in this work -- is that the angular size of the source is much smaller than that of the dark matter substructure.  We approximate the source by a point.  However, for high magnification images, the area of the triangle on the source plane may be smaller than observed source sizes, as discussed in Sec.~\ref{sec:caveats}.

For each macromodel and each substructure realization, we sample the source plane 50,000 times, creating $\sim10,000$ mock four image lens systems.  We confine our investigation to four image systems, since those have a sufficient number of observables and have been proved to be most useful for substructure studies.

\section{Comparisons to Observational Data}
\label{sec:compare}

\begin{figure}
\centering
\plotone{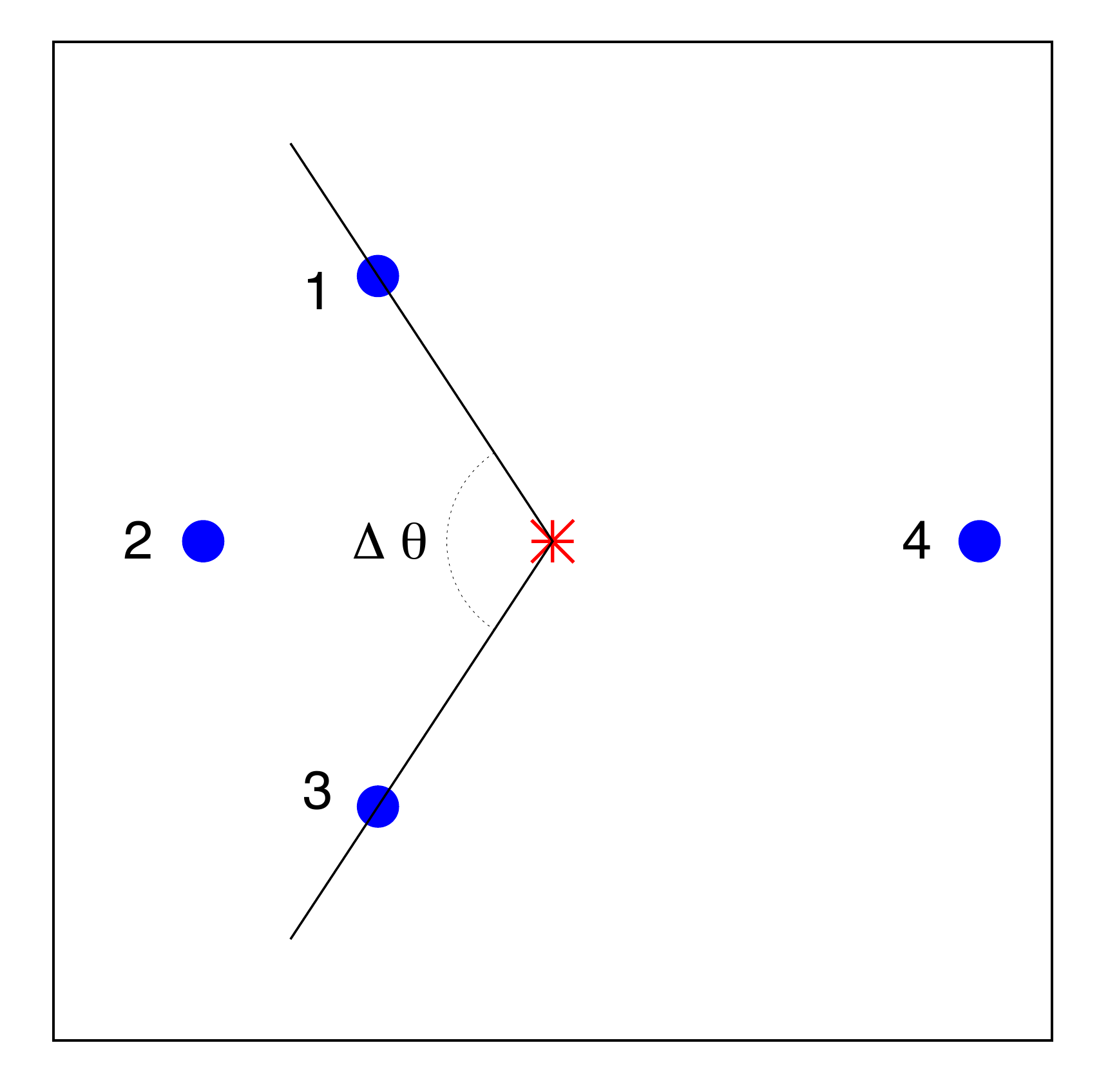}
\caption{Sample mock four-image lens, with the lensed images (points and labeled by number), the lens center (star), and the opening angle for the cusp relation labeled. \label{fig:cartoon}}
\end{figure}

A parameter that can be used to characterize deviations from a simple, smooth lens is 
the signed sum of the image fluxes from any three adjacent images in a four-image system.  
We work with a scaled version of such a sum, 
\begin{equation}
\label{eq:rcusp}
R_{\rm cusp} \equiv \frac{| \mu_1 + \mu_2 + \mu_3|}{|\mu_1| + |\mu_2| + |\mu_3|} = \frac{| F_1 + F_2 + F_3|}{|F_1| + |F_2| + |F_3|},
\end{equation}
where $\mu_{i}$ is the magnification and $F_{i}$ is the observed flux, and 
both are signed quantities which indicate the image parities.  In practice, 
determining the image parities is unnecessary;  in a triplet of images, the outer two 
images have one image parity and the middle image has the opposite parity.  
The ideal cusp relation has $R_{\rm cusp} = 0$ in cases where the lens potential is 
smooth and the opening angle $\Delta \theta$ spanned by the images from a vertex 
at the lens center approaches zero.  An illustrative example of a mock lens system, 
providing a clear definition of $\Delta \theta$, is shown in Fig.~\ref{fig:cartoon}.   
In practice, we take the sum in Eq.~(\ref{eq:rcusp}) over the three images that 
yield the smallest opening angle.  The cusp relation is the measure used in many anomalous 
flux ratio studies \citep[e.g.,][]{maccio_etal06,amara_etal06,metcalf_amara10}.

Table \ref{tab:obs} lists $R_{\rm cusp}$ values for systems with a single lens galaxy and four distinct, point-like images, observed in the radio or mid-infrared (noted by "IR").  This excludes B1608+656 \citep{koopmans_fassnacht99,fassnacht_etal02}, B1127+385 \citep{koopmans_etal99}, and B1359+154 \citep{myers_etal99,rusin_etal01} which have multiple lensing galaxies.  Limiting ourselves to distinct images also excludes MG 2016+112 \citep{garrett_etal94,garrett_etal96,koopmans_etal02}, MG 0751+2716 \citep{lehar_etal97,tonry_kochanek99}, and B1938+666 \citep{king_etal97,tonry_kochanek00}, in which 
merging pairs or arcs are seen.  We specifically include the lens systems which have observable satellite galaxies near the lens, B2045+265 and MG 0414+053, which each have one.  Not all of the systems in Table \ref{tab:obs} are nearly cusp catastrophes -- i.e., a cusp configuration where $\Delta \theta \rightarrow 0$ --
but in all cases $R_{\rm cusp}$ is calculated from the 3 images with the smallest opening angle.  

Radio data has been traditionally used in lensing studies of substructure as it is free of microlensing contamination from stars in lensing galaxies.  This is because the size of the quasar source in the radio ($\sim$$[1-10]$ parsecs) is larger than the scale size of perturbations from stars \citep[e.g.,][]{mao_schneider98} and from subsolar mass dark matter halos \citep{chen_koushiappas10}.  The size of the source relative to the perturbations from DM substructure, on the other hand, is very small. Therefore, in our simulations we approximate the source as a point.  Mid-infrared observations are probably the best flux measurements for substructure lensing, as the flux measurements are additionally free from extinction effects, as well as from microlensing effects \citep{minezaki_etal04,chiba_etal05}.  Only five lens systems, however, have been observed in the mid-IR (see Table \ref{tab:obs}).

\begin{table}[h]
\begin{center}
\begin{tabular}{ll|c|c|c}
\hline
& System Name & $\Delta \theta$($\degr$) & $R_{\rm cusp}$ & Reference\\
\hline
\hline
1.)& B2045+265 & 34.9 & 0.501 & \citet{fassnacht_etal99},\\
&&&&\citet{koopmans_etal03} \\

2.) & B0712+472 & 76.9 & 0.255 & \citet{jackson_etal98},\\
&&&&\citet{koopmans_etal03} \\

3.)& B1422+231 & 77.0 & 0.187 & \citet{patnaik_etal99},\\
&&&&\citet{koopmans_etal03}\\
&& & 0.251 & \citet[][IR]{chiba_etal05}\\

4.) &  MG 0414+053 & 101.5 & 0.227 & \citet{katz_etal97} \\
 &&& 0.204 & \citet[][IR]{minezaki_etal09} \\

5.) &B1555+375 & 102.6 & 0.417 & \citet{marlow_etal99},\\
&&&&\citet{koopmans_etal03} \\

6.)&B0128+437 & 123.3 & 0.01 & \citet{phillips_etal00} \\

7.) &PG1115+080 & 127.5 & 0.110 & \citet[][IR]{chiba_etal05}\\

8.)& B1933+503 & 143.0 & 0.39 & \citet{cohn_etal01}\\

9.)& Q2237+030 & 146.3 & 0.357 & \citet{falco_etal96} \\
& &  & 0.270 & \citet[][IR]{minezaki_etal09} \\

10.)& H1413+117 & 160.4 & 0.22 & \citet[IR]{macleod_etal09} \\

\hline
\end{tabular} 
\caption{Observed Lens Systems}
\label{tab:obs}
\end{center}
\end{table}

\begin{figure}
\centering
\plotone{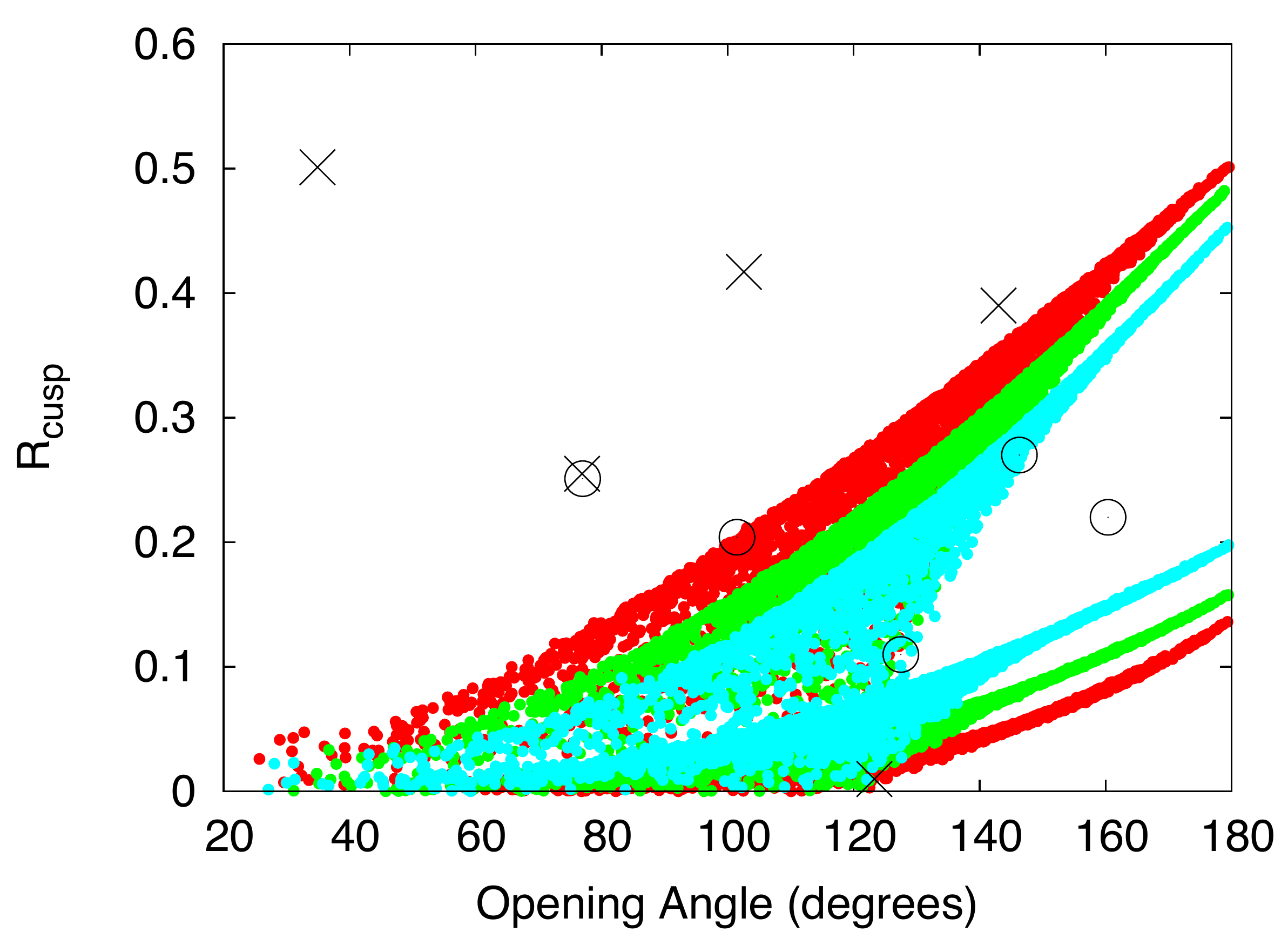}
\caption{Values of $R_{\rm cusp}$ for macromodels M1 (blue dots), M2 (red dots), and M3 (green dots) without including substructure.  The observed $R_{\rm cusp}$ values from Table \ref{tab:obs} are shown as black crosses for radio data and as circles for IR data, where mid-IR values are used in place of radio values when both are available. \label{fig:rcusp_nosubs}}
\end{figure}

\begin{figure*}
\centering
\resizebox{2.in}{!}	{\includegraphics{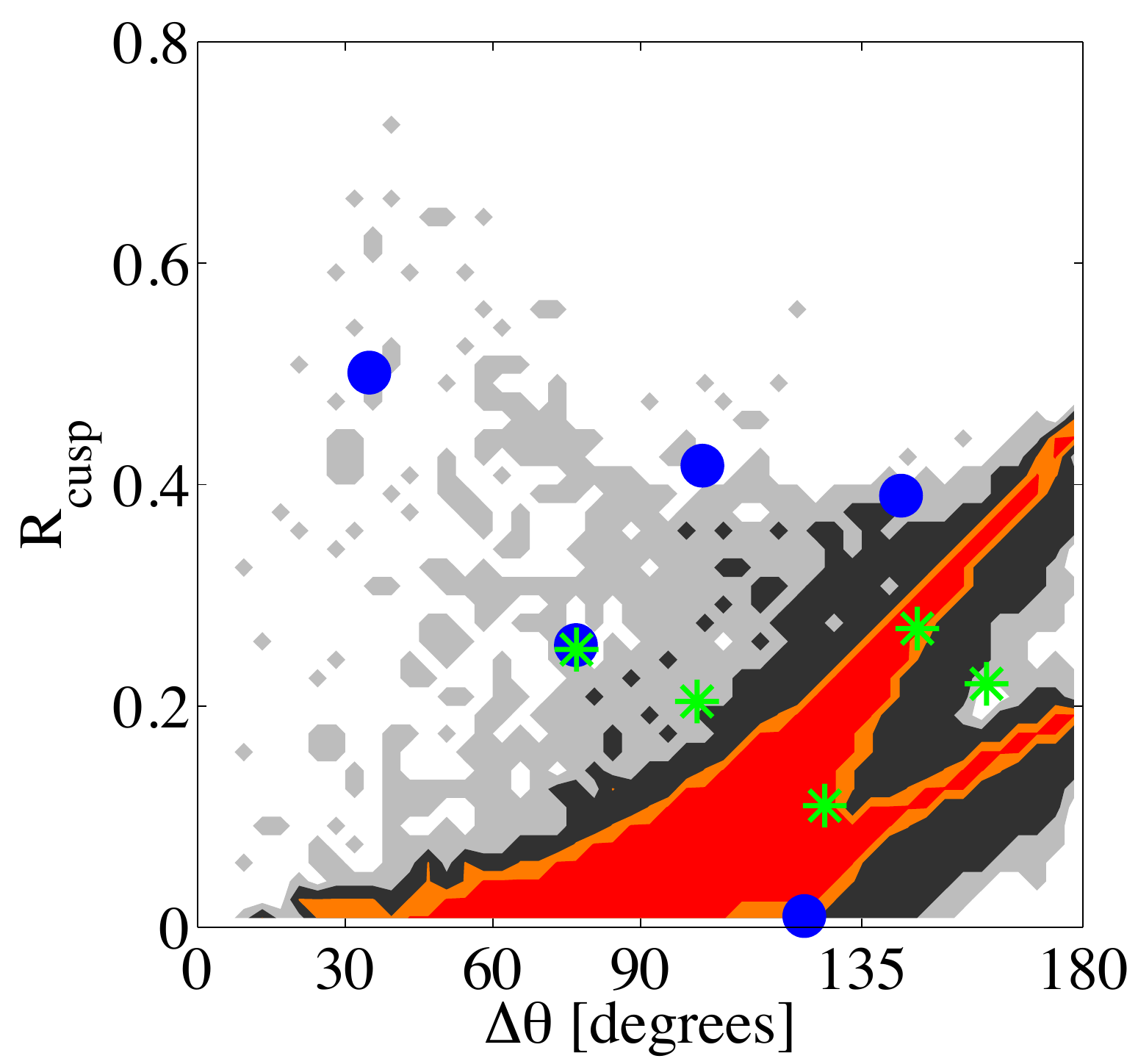}}
\resizebox{2.in}{!}	{\includegraphics{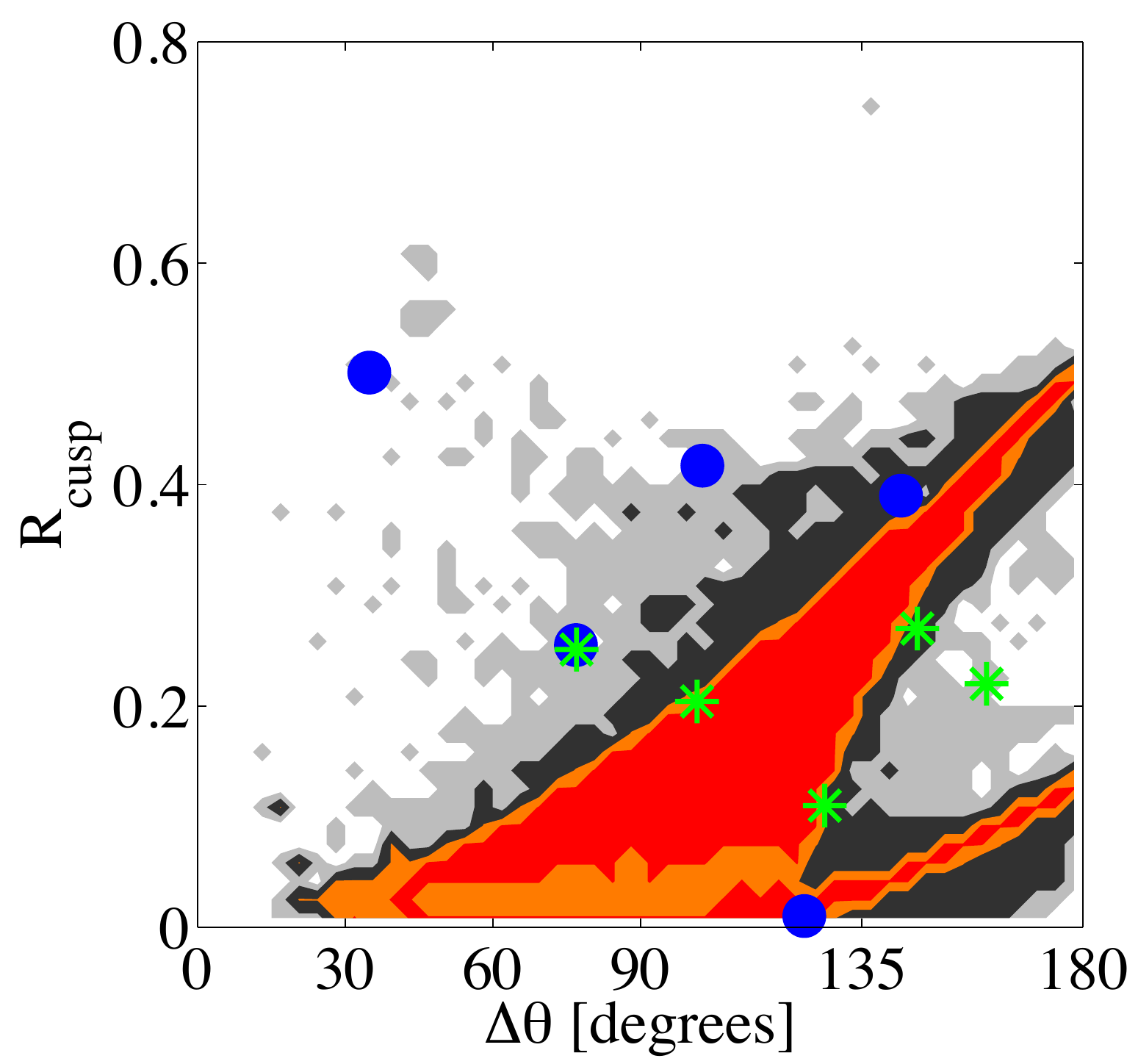}}
\resizebox{2.in}{!}	{\includegraphics{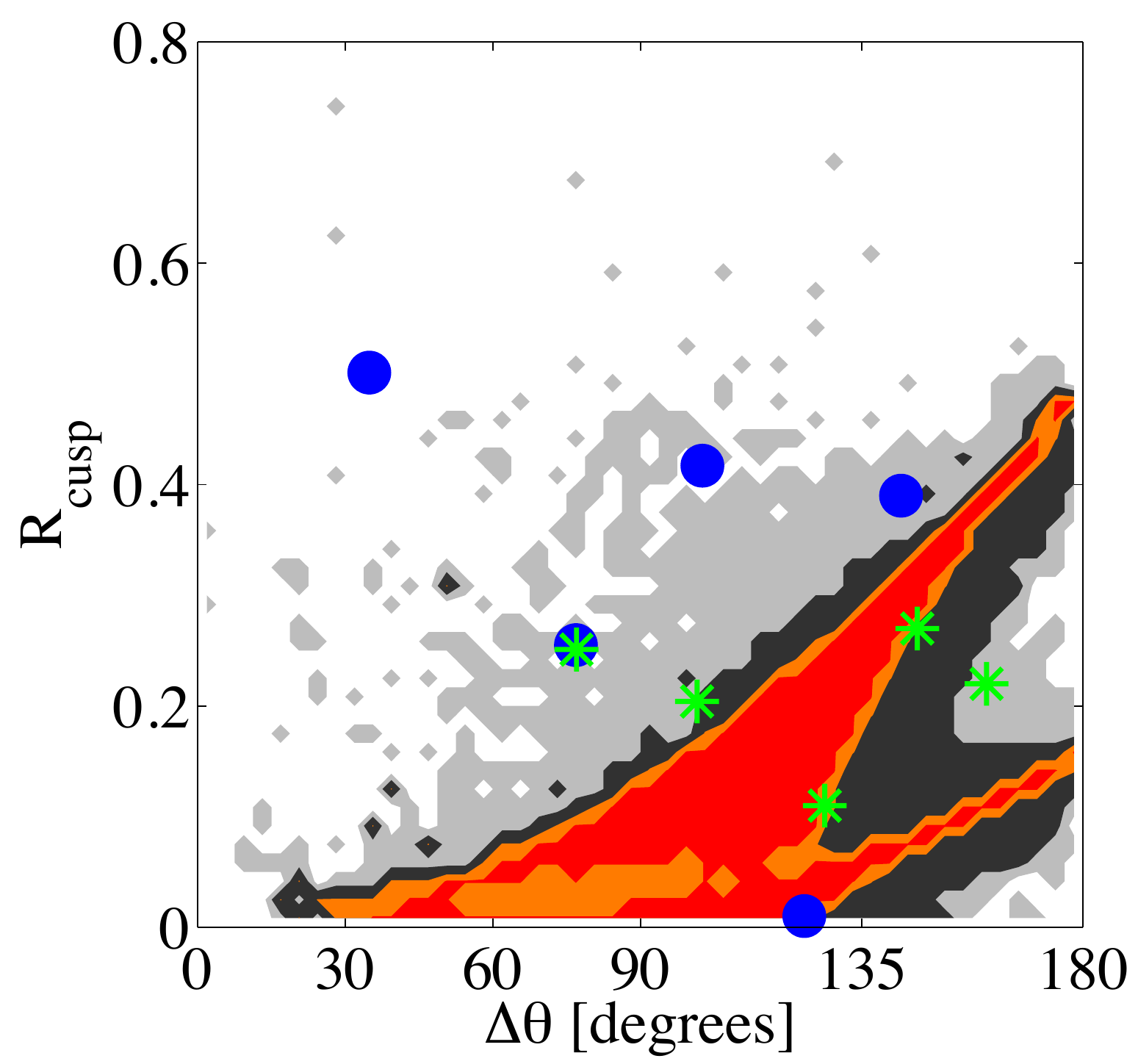}}
\caption{Values of $R_{\rm cusp}$ for Models 1 (left panel), 2 (center panel), and 3 (right panel) including all substructure realizations (light and dark gray areas) and overplotted with no substructure realizations (orange and red areas).  Red and dark gray areas show where 99\% of the values lie, while light gray and orange show where the remaining 1\% of the values are found.  The observed $R_{\rm cusp}$ values from Table \ref{tab:obs} are shown in blue dots for radio data and green asterisks for mid-IR,  where mid-IR values are used in place of radio values when both are available. 
\label{fig:rcusp_m}}
\end{figure*}

\begin{table*}[ht]
\begin{center}
\begin{tabular}{c|c|c|c|c|c}
\hline
$\Delta \theta$ & Macromodel &$R_{\rm cusp}$ (RS)\footnote{The largest $R_{\rm cusp}$ value in models with no substructure.} & P($R_{\rm cusp} > RS$) & $R_{\rm cusp}$ (RO)\footnote{The largest observed value of $R_{\rm cusp}$.} & P($R_{\rm cusp} > RO$)  \\ 
\hline
\hline
$30-40\degr$ &M1 & 0.1  & 0.110 & 0.5 & 0.042\\ 
 &M2 &           & 0.048 &          & 0.001 \\
&M3 &           & 0.042 &          & 0.001 \\ 
\hline
$70-80\degr$ &M1 &  0.15 & 0.0198 & 0.25 & 0.0121\\ 
 &M2 &           & 0.0320 &          & 0.0138 \\ 
 &M3 &           & 0.0148 &          & 0.0062 \\ 
\hline
$100-110\degr$ &M1 & 0.25  & 0.0070 &  0.4 & 0.0008\\ 
 &M2 &           & 0.0171&           &  0.0028\\ 
 &M3 &           & 0.0052 &           &  0.0014\\ 
\hline
\end{tabular} 
\caption{Cumulative Probability of Cusp Relation Values}
\label{tab:results}
\end{center}
\end{table*}

\begin{figure}
\centering
\resizebox{3.15in}{!}	{\includegraphics{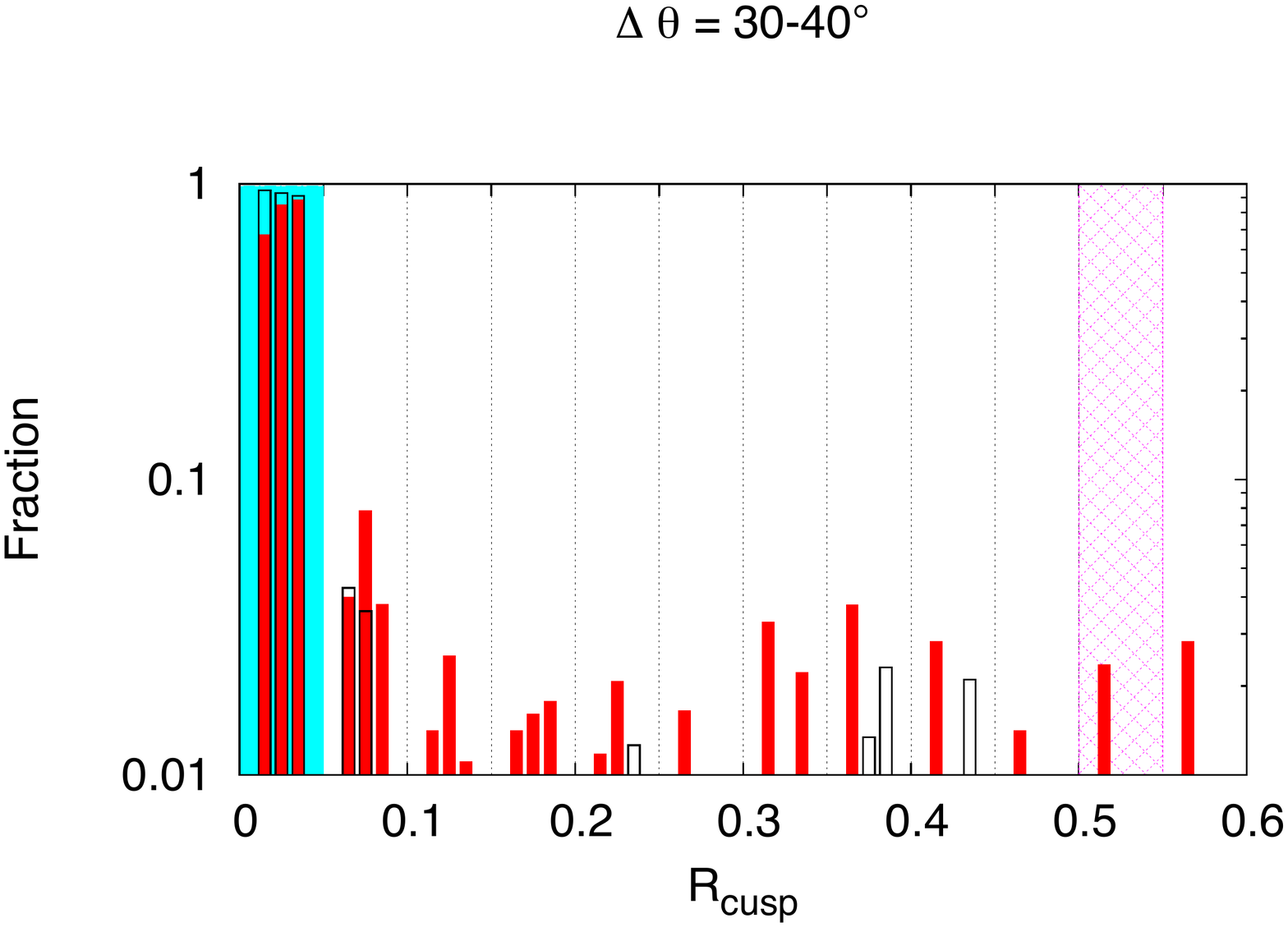}}\\
\resizebox{3.15in}{!}	{\includegraphics{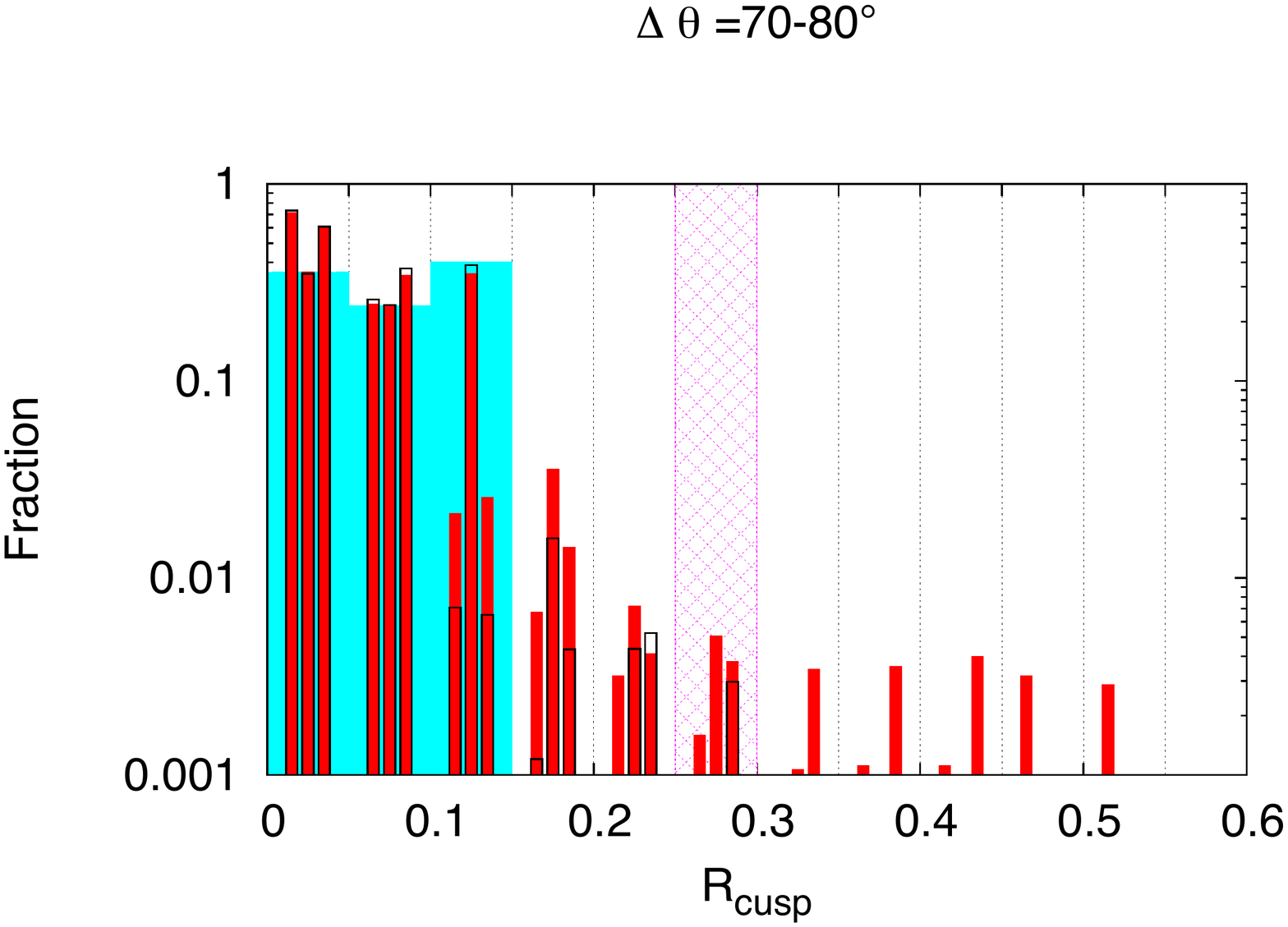}}\\
\resizebox{3.15in}{!}	{\includegraphics{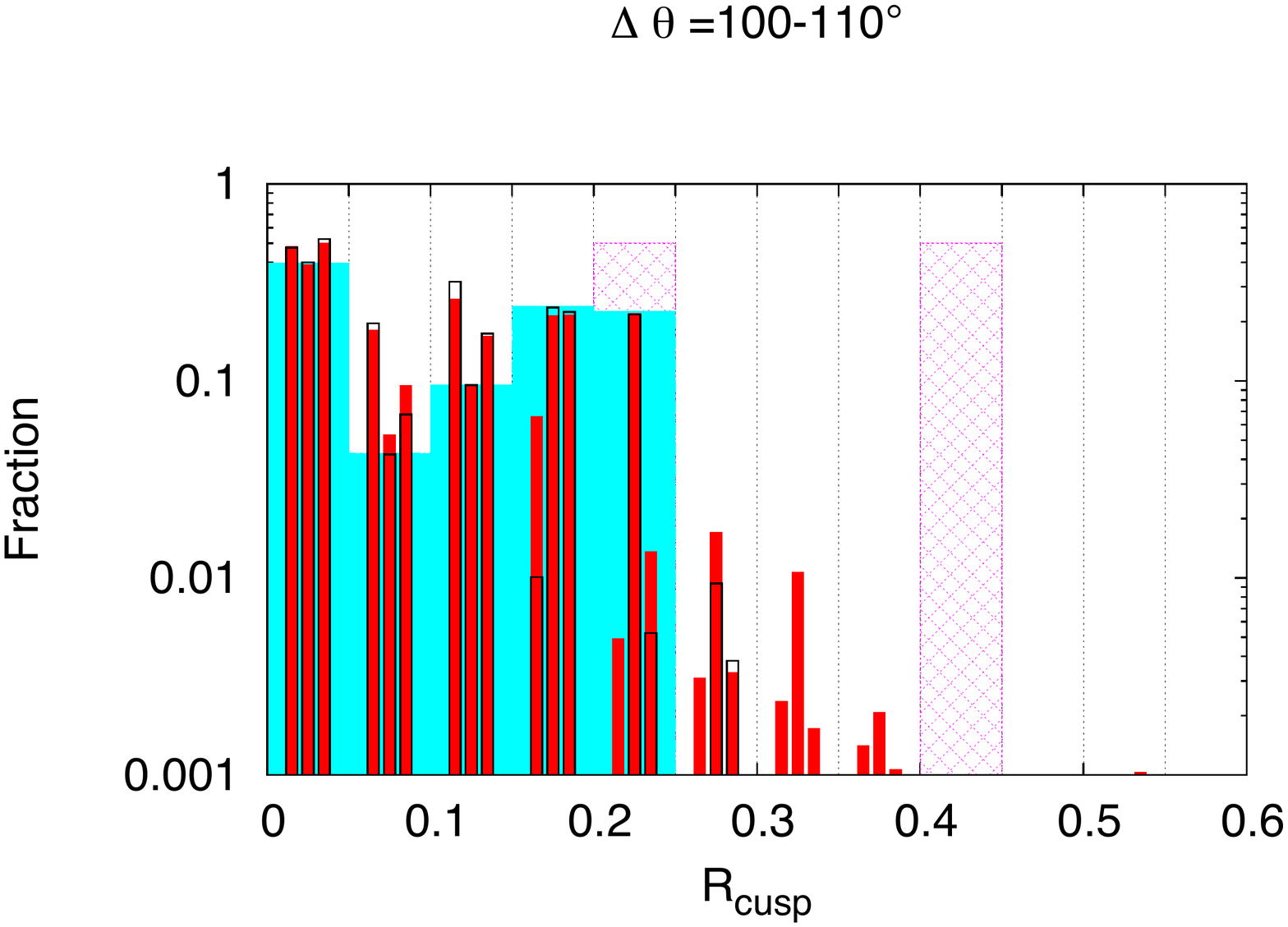}}
\caption{Histogram of $R_{\rm cusp}$ values for the substructure realizations where the host halo concentration is in the lowest 16 percent (red bars) and in the highest 16 percent (black lines).  At each $R_{\rm cusp}$ bin, the bars show the results for M1, M2, and M3 from left to right.  The cyan colored bars show the results using the M2 macromodel parameters and {\it no} substructure.  The cross-hatched magenta shows the observed results.  From top to bottom, the panels show opening angles of 30-40$\degr$, 70-80$\degr$, and 100-110$\degr$.  
\label{fig:conc}}
\end{figure}

\begin{figure*}
\centering
\resizebox{2.12in}{!}	{\includegraphics{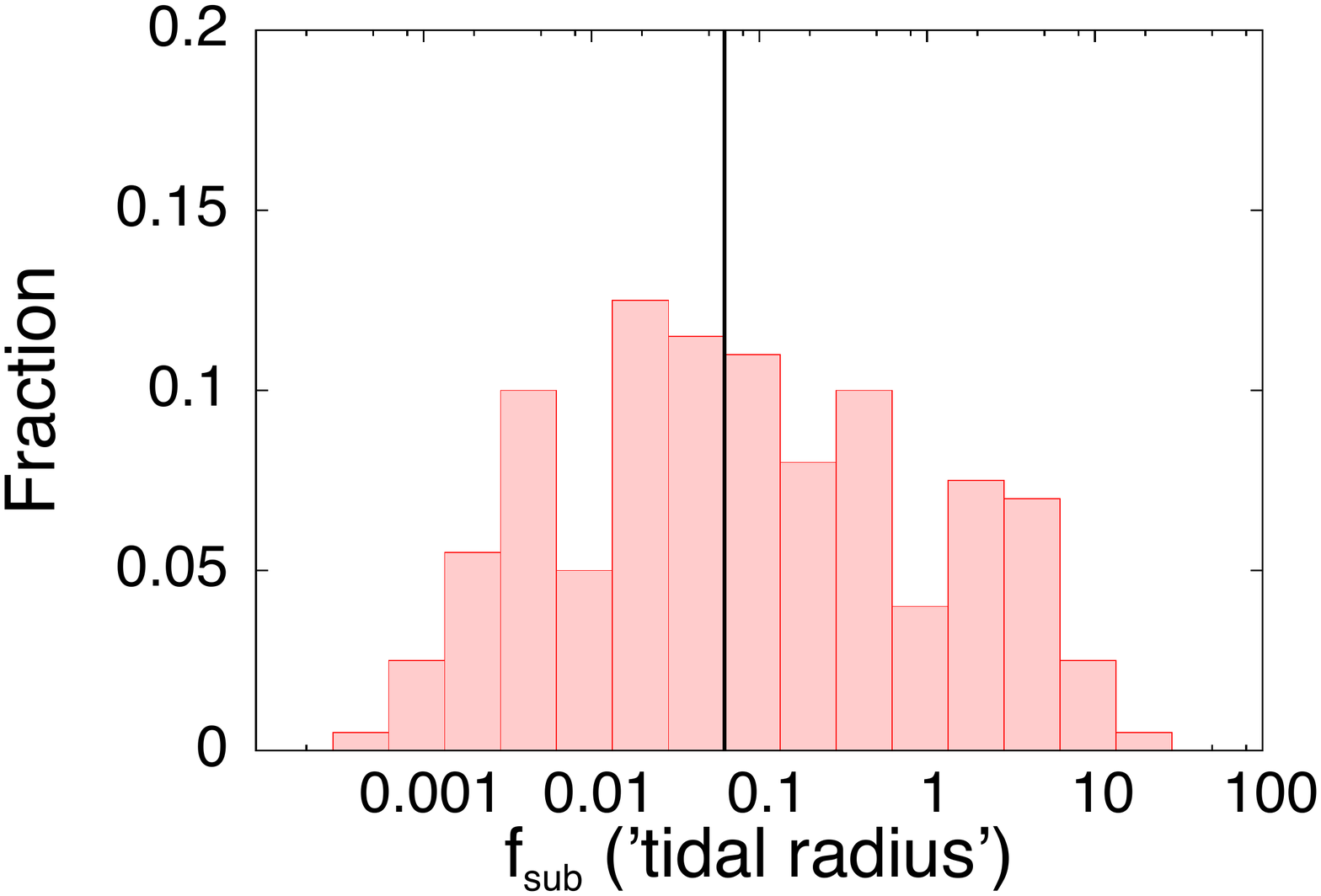}}
\resizebox{2.12in}{!}	{\includegraphics{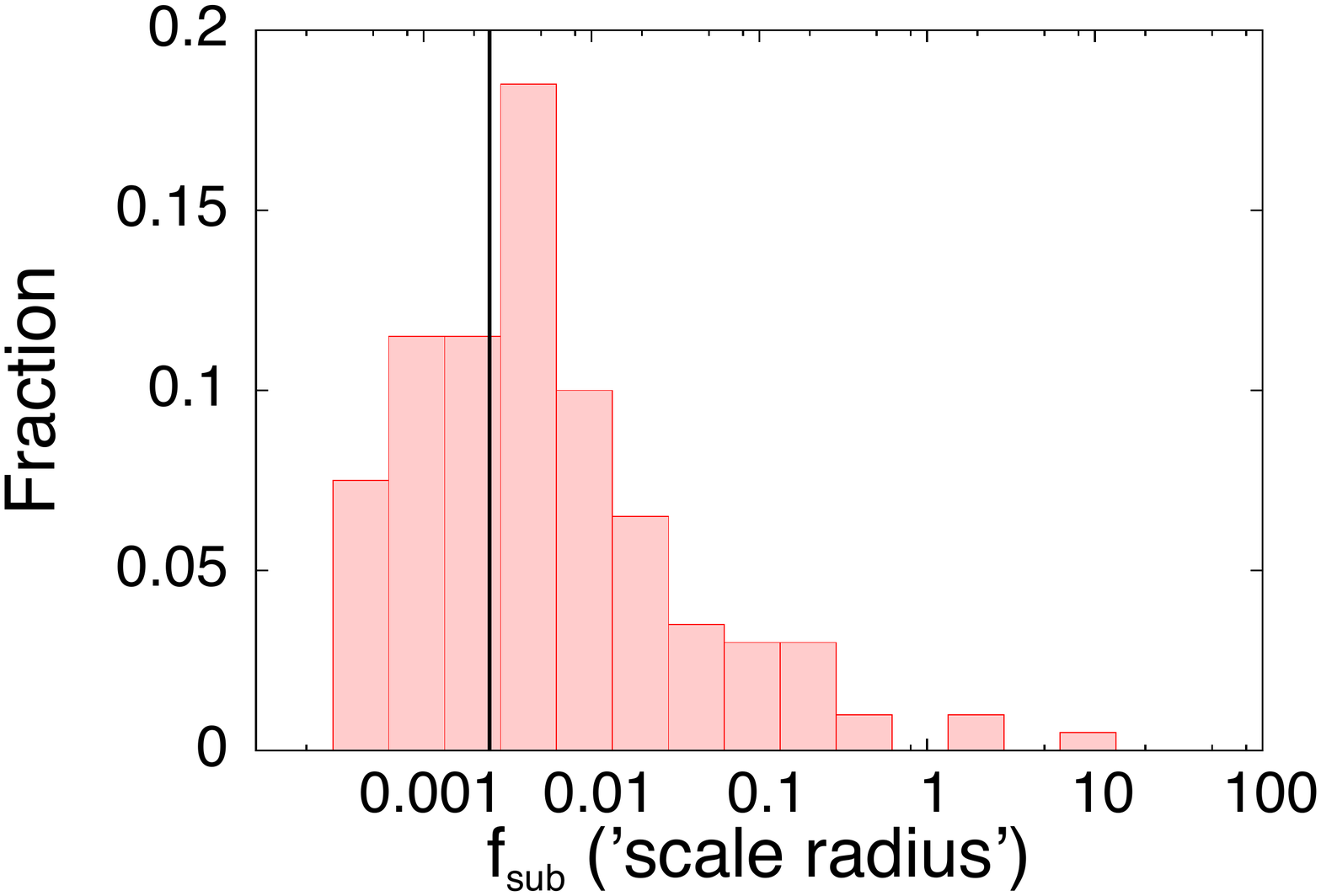}}
\resizebox{2.12in}{!}	{\includegraphics{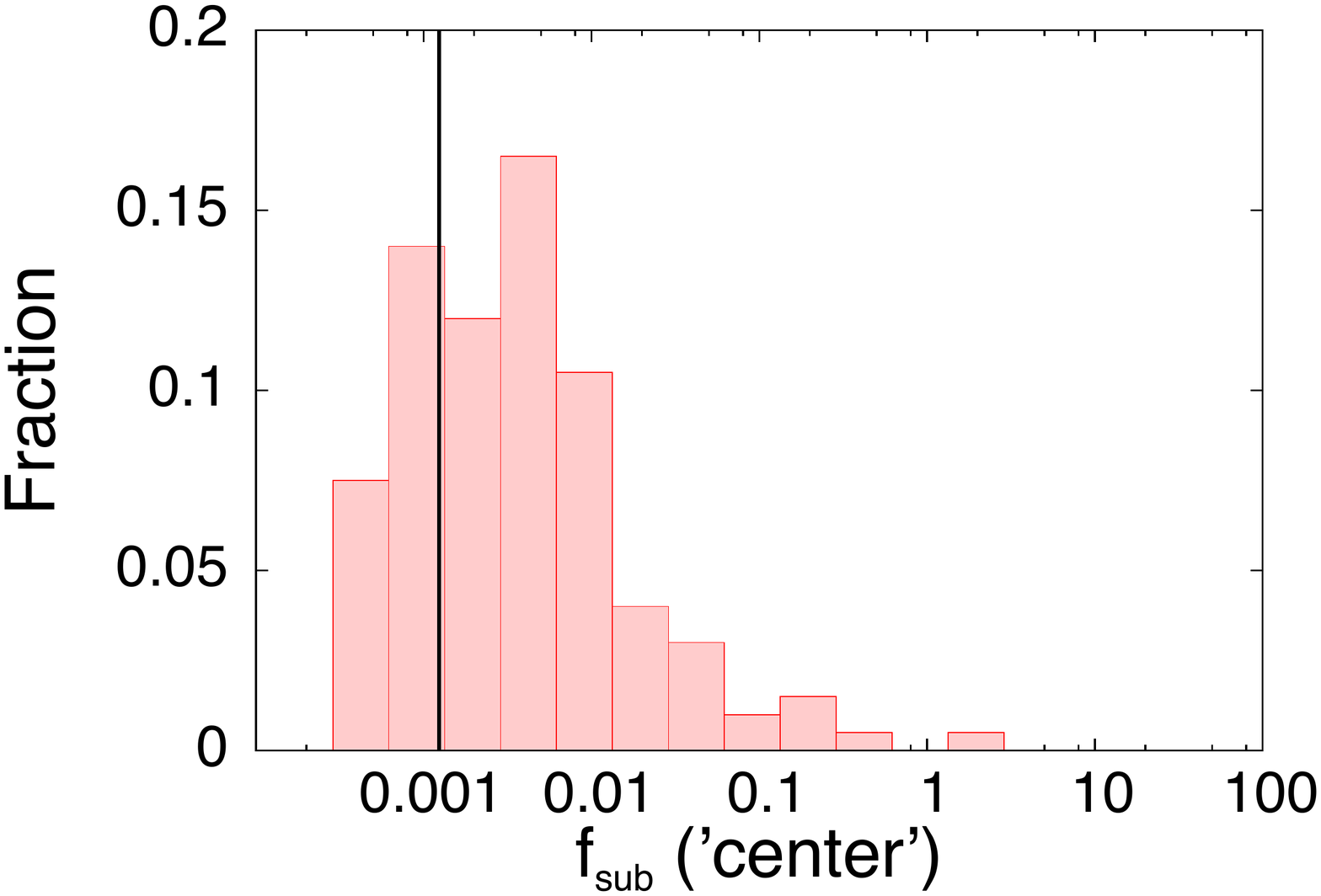}}

\resizebox{2.12in}{!}	{\includegraphics{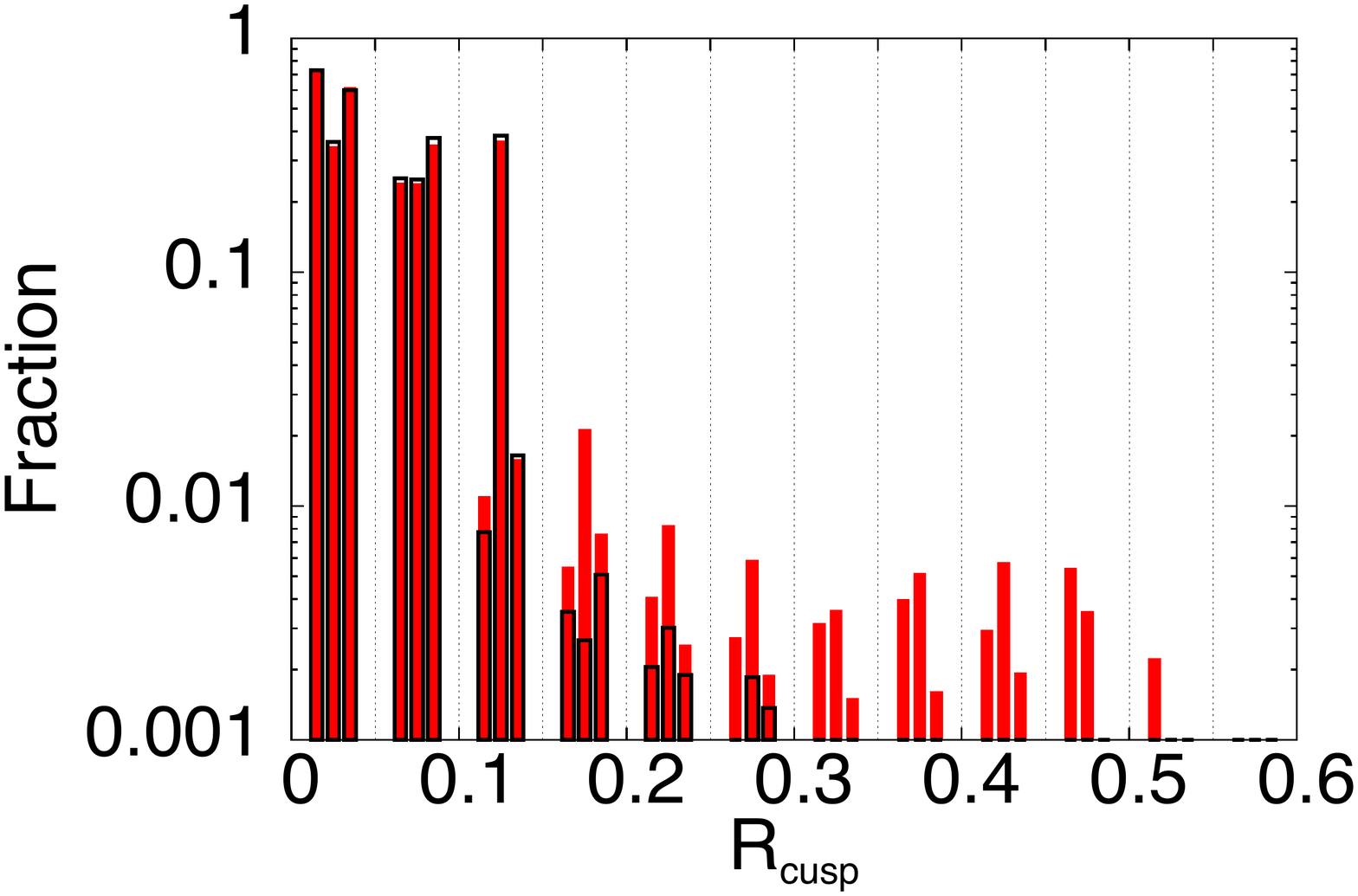}}
\resizebox{2.12in}{!}	{\includegraphics{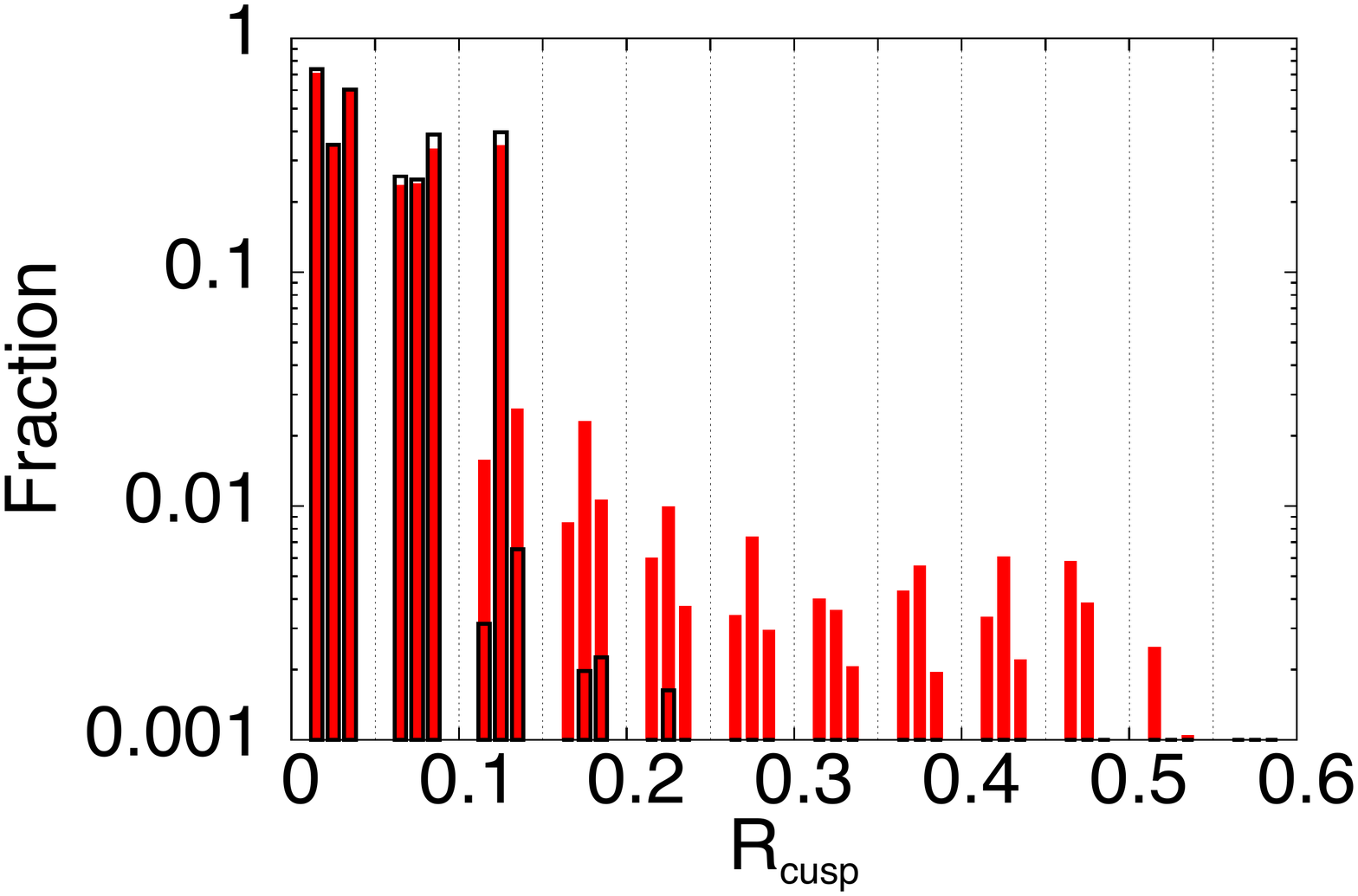}}
\resizebox{2.12in}{!}	{\includegraphics{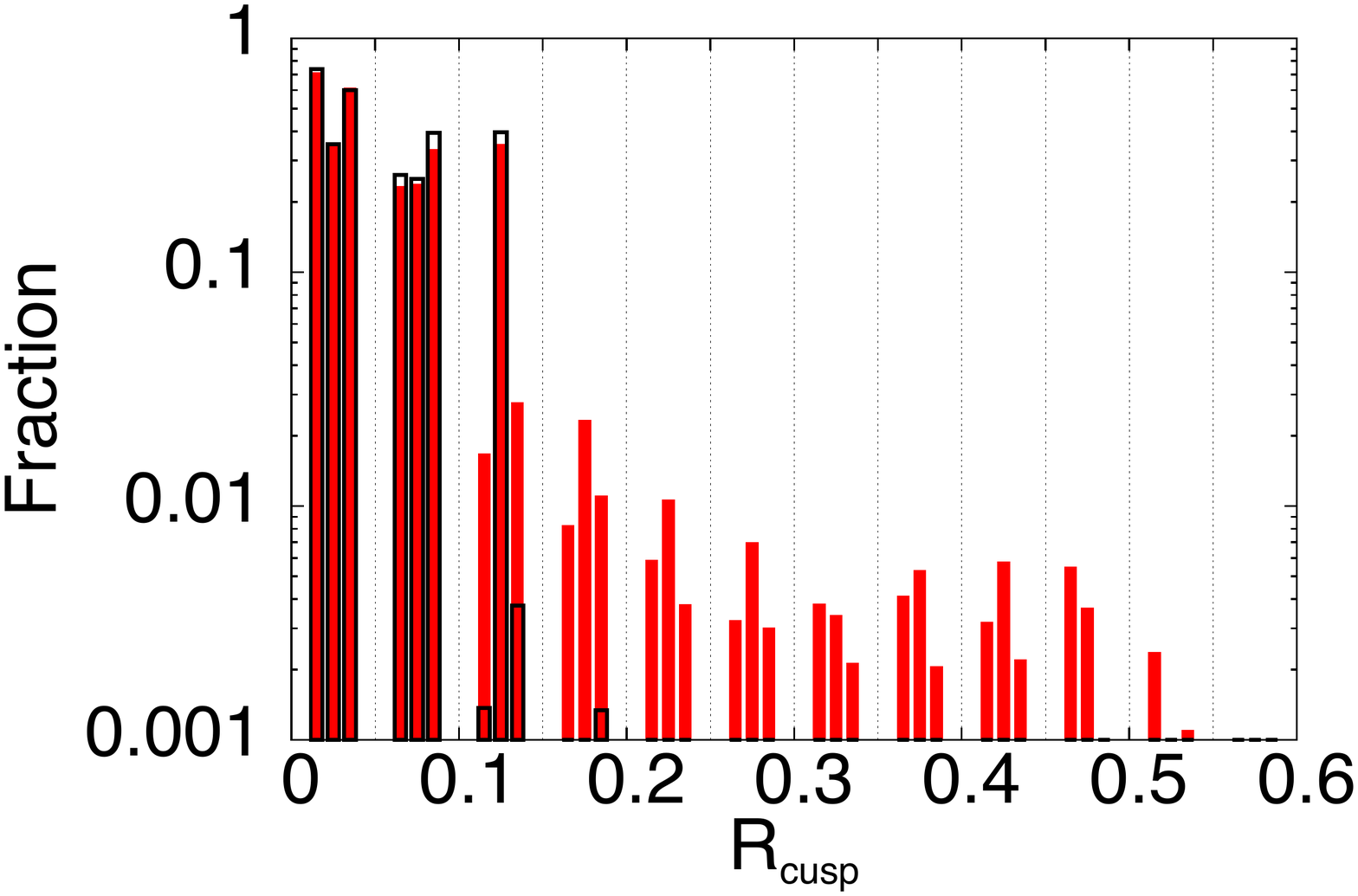}}
\caption{{\it Top:}  Distribution of $f_{\rm sub}$ values for three different criteria.  See text for additional details.  The black vertical lines show the sample median.  Some substructure realizations have $f_{\rm sub}=0$ and are not shown in the panels.  {\it Bottom:}  The corresponding distribution of  $R_{\rm cusp}$ values for $\Delta \theta = 70-80\degr$.   The substructure realizations where $f_{\rm sub}$ is greater than the median are shown with filled red bars, and realizations where $f_{\rm sub}$ is less than the median are shown with black lines.  In each $R_{\rm cusp}$ bin, the results from M1, M2, and M3 are shown from left to right.   
\label{fig:cuts}}
\end{figure*}

\begin{figure}
\centering
\resizebox{3.5in}{!}	{\includegraphics{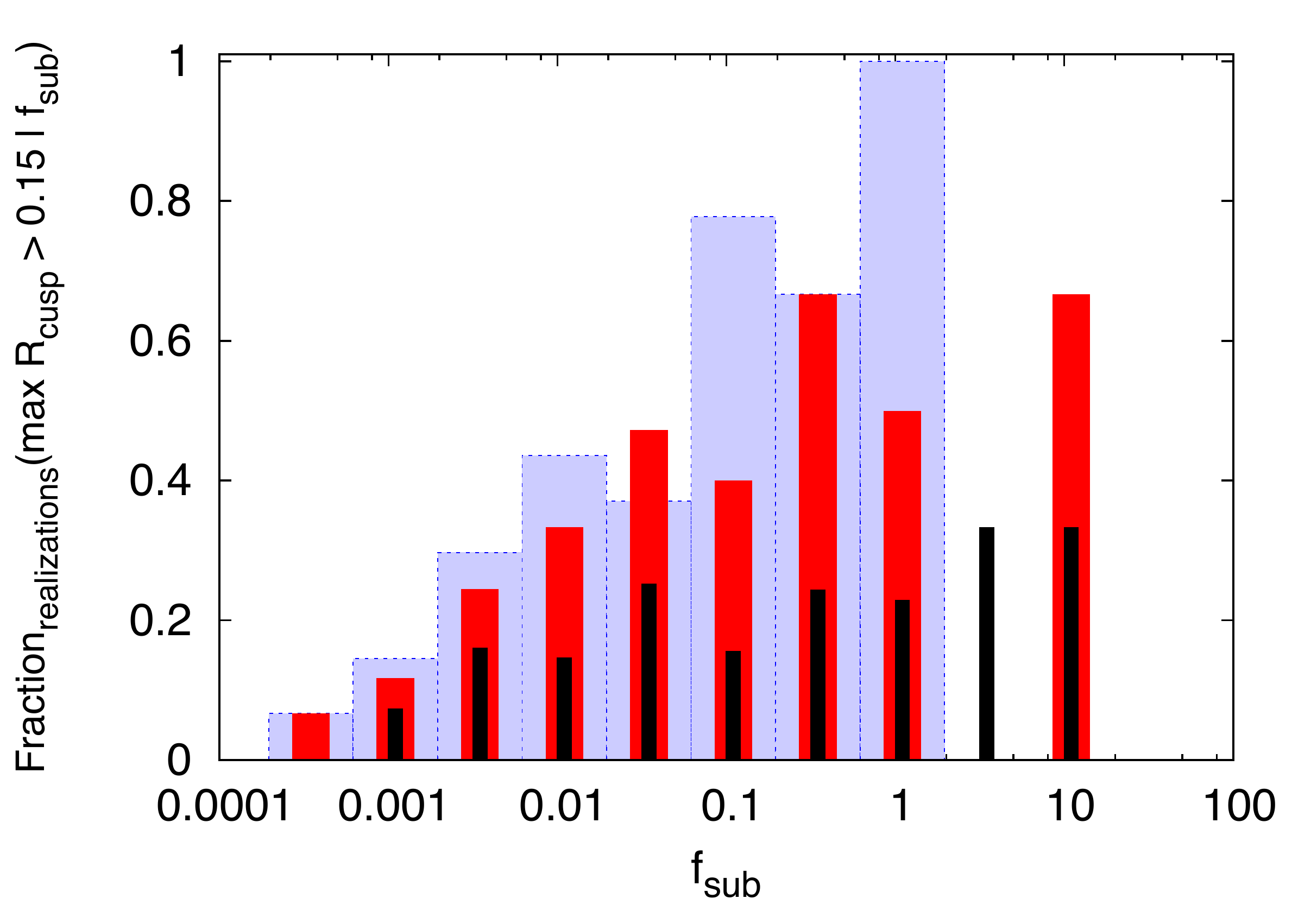}}
\resizebox{3.5in}{!}	{\includegraphics{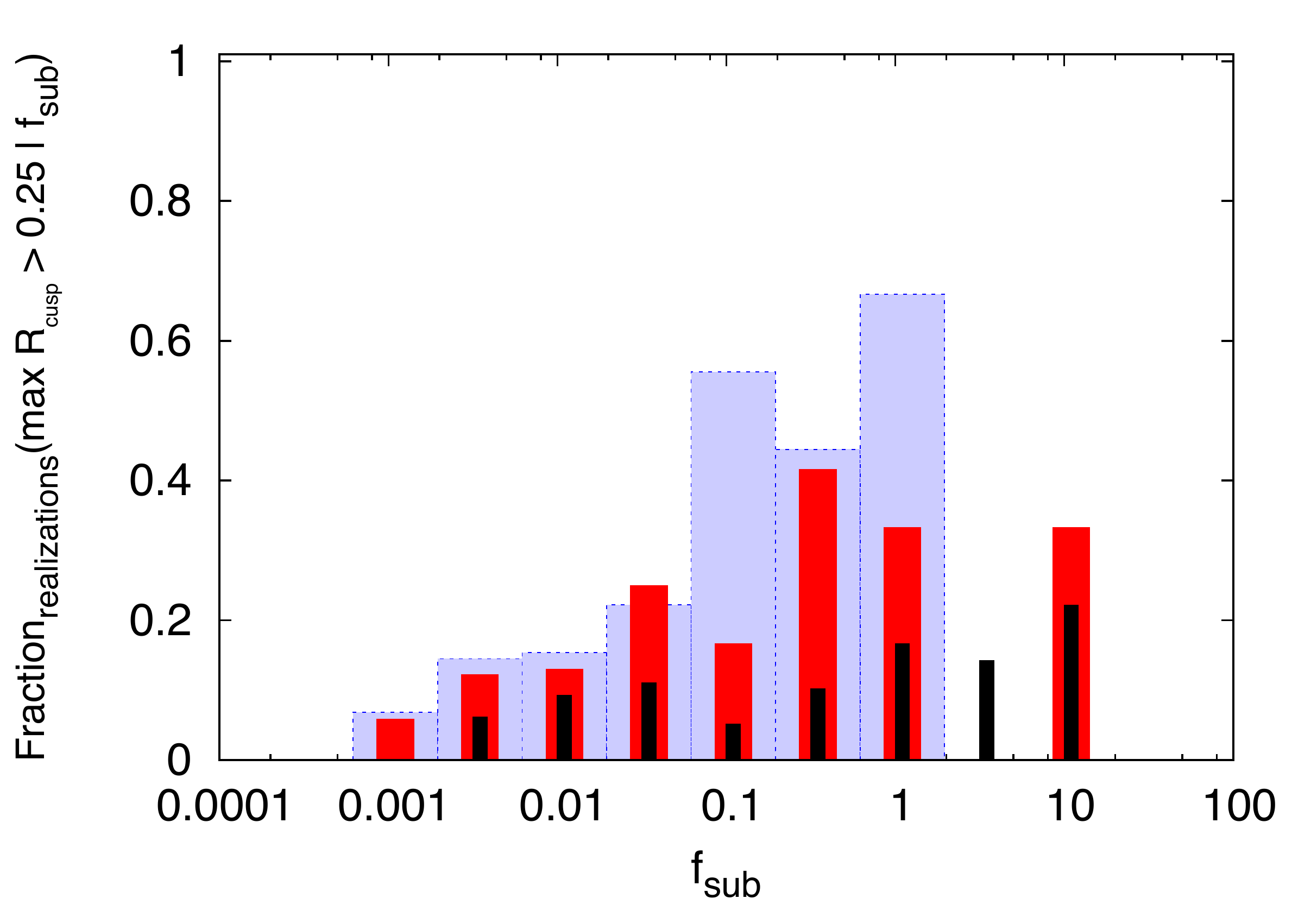}}
\caption{Fraction of realizations in each bin of $f_{\rm sub}$ which also contain $R_{\rm cusp} > R_{\rm cut}$ values.  Three different criteria for $f_{\rm sub}$ are considered:  including substructures whose tidal radii fall within $R=2b$ (black bars);  including substructures whose scale radius fall within $R=2b$  (red bars); and including substructures whose halo centers fall within $R=2b$ (blue bars).  Opening angles of  70-80$\degr$ are shown and results using macromodels M1, M2, and M3 are combined.  The threshold values, $R_{\rm cut}=0.15$ (top) and $R_{\rm cut}=0.25$ (bottom) are taken from Table \ref{tab:results} and represent the $R_{\rm cusp}$ value that is greater than provided by smooth models and $R_{\rm cusp}$ value that is greater than the observed values, respectively. ~~~~~~
\label{fig:T345}}
\end{figure}

\begin{figure*}
\centering
\resizebox{3.2in}{!}	{\includegraphics{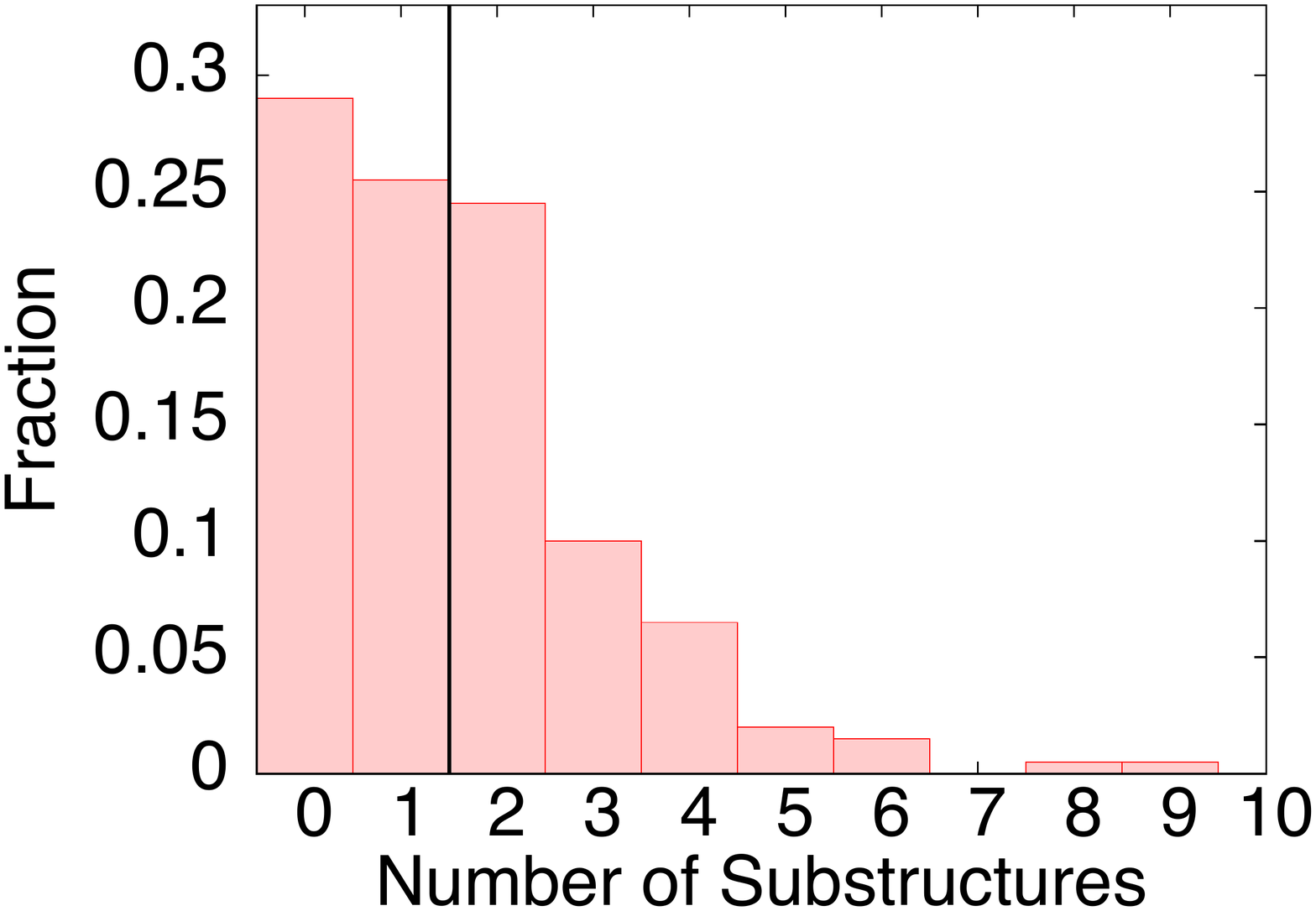}}
\resizebox{3.2in}{!}	{\includegraphics{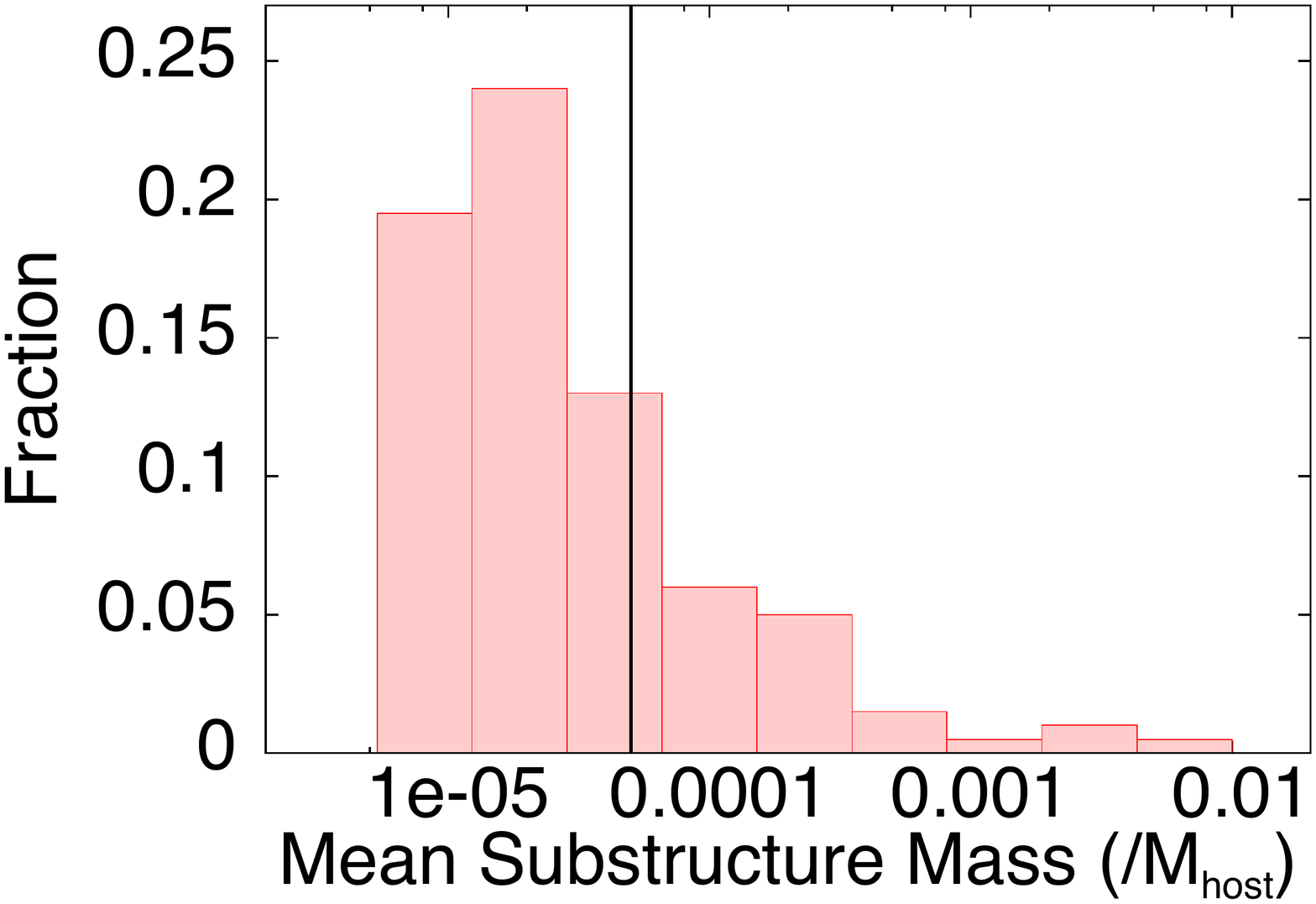}}

\resizebox{3.2in}{!}	{\includegraphics{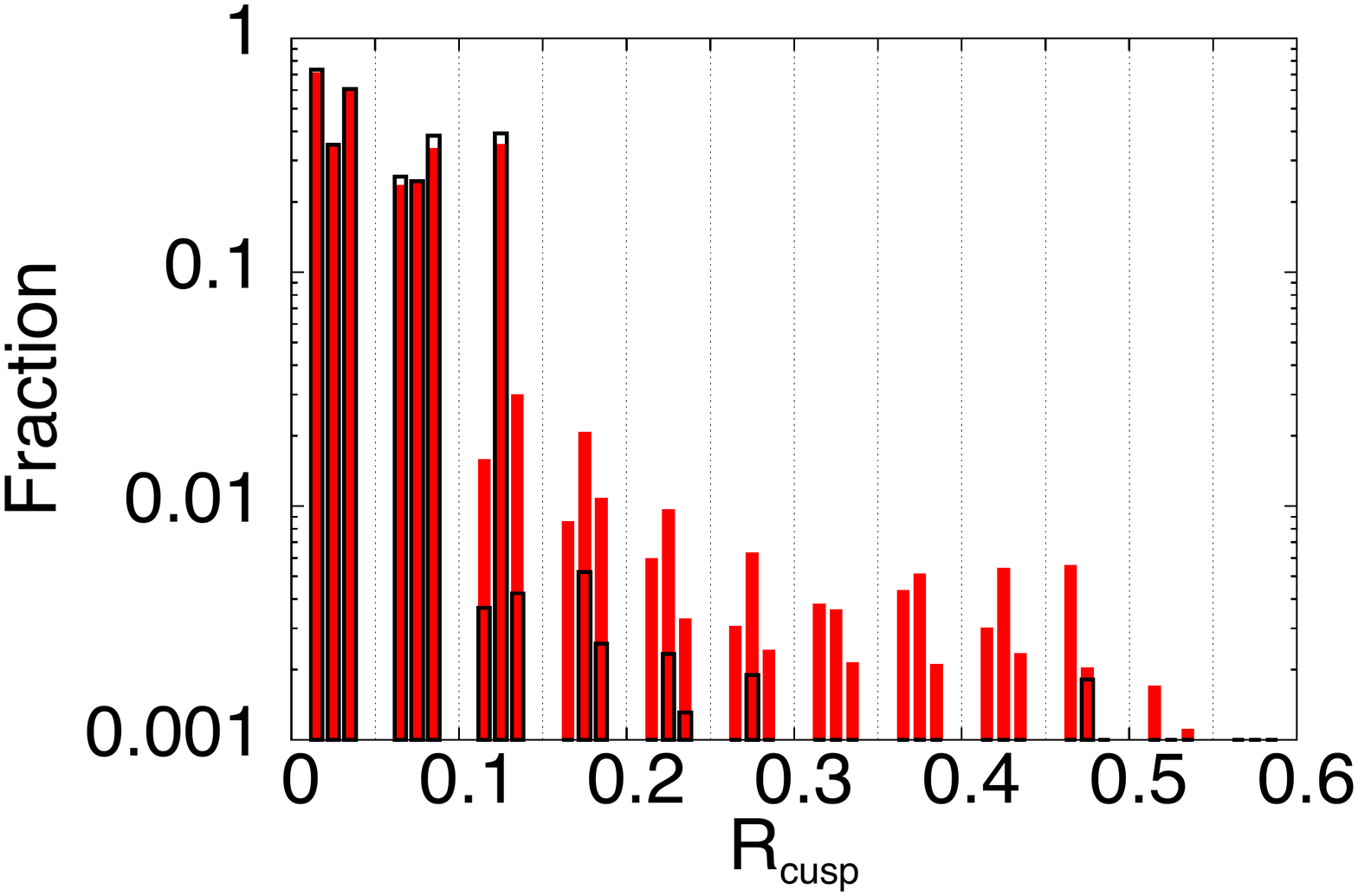}}
\resizebox{3.2in}{!}	{\includegraphics{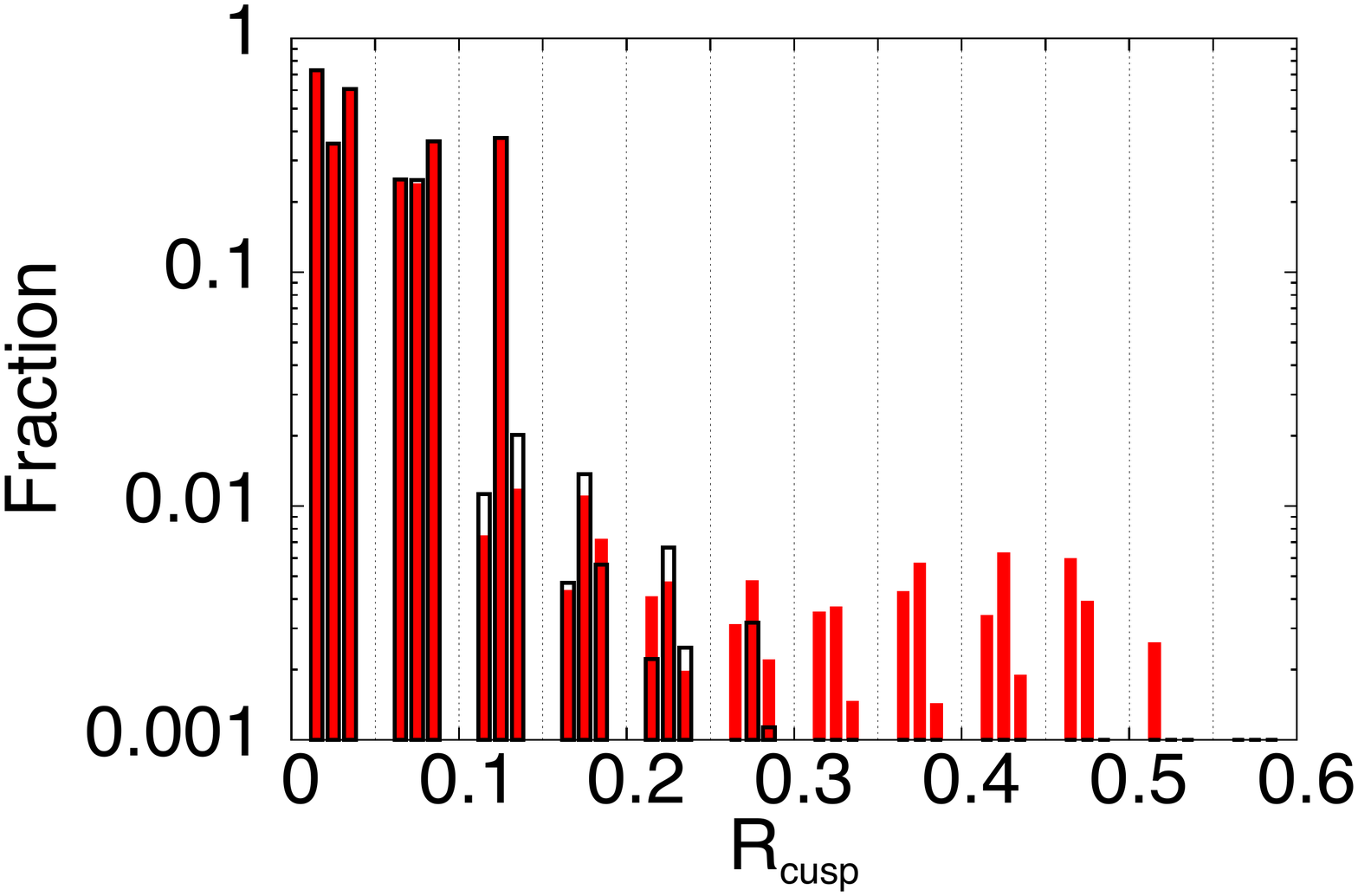}}
\caption{{\it Top:}  The number of subhalos whose centers fall within $R=2b$ (left), and the mean mass for those subhalos (right). The black vertical lines show the sample median.  {\it Bottom:}  The corresponding distribution of  $R_{\rm cusp}$ values for $\Delta \theta = 70-80\degr$.   The substructure realizations where number of mass is greater than the median is shown in red bars, and realizations where it is less than the median in shown in black lines.  In each $R_{\rm cusp}$ bin, the results from M1, M2, and M3 are shown from left to right.   
\label{fig:cuts2}}
\end{figure*}

Fig. \ref{fig:rcusp_nosubs} shows the values of $R_{\rm cusp}$ for macromodels M1 (blue dots), M2 (red dots), and M3 (green dots) without including the effects of substructure.  The observed $R_{\rm cusp}$ values from Table \ref{tab:obs} are shown as black crosses and circles, where mid-IR values are used in place of radio values where both are available.   Small opening angles correspond to small cusp relation values, as is expected.  At larger opening angles, for each model the values of $R_{\rm cusp}$ bifurcate, corresponding to source positions along the minor axis (smaller $R_{\rm cusp}$ values) or the major axis (larger $R_{\rm cusp}$ values) of the lensing halo.  As expected, values of $R_{\rm cusp}$ depend on the macromodel parameters \citep{keeton_etal03}.  At the smallest of opening angles, $R_{\rm cusp} \rightarrow 0$, regardless of macromodel.  At opening angles, $\lesssim 90\degr$, the observed values of $R_{\rm cusp}$ clearly deviate from what may be expected from smooth models.  

The effect of substructure on the cusp relation is shown in Fig. \ref{fig:rcusp_m}.  The no substructure results from Fig. \ref{fig:rcusp_nosubs} are reproduced here in red and orange shaded areas.   The borders of the colored areas are contours of equal probability, and the red and orange patches encompass the areas where 99\% and 100\% of the cusp relation values are found.  These areas are well contained within a small portion of the $R_{\rm cusp}-\Delta \theta$ parameter space.  Including the effect of substructure gives values which are shown by the dark gray and light gray shaded areas.  The dark gray area, which encompasses the area where 99\% of the values are found, overlaps significantly with the region in which the no substructure values lie.  And where the substructure-included values do extend beyond this area, they mostly lie fairly close to the no substructure values.  In these cases, the deviations from the no substructure case are not due to small perturbations to individual images but perturbations that effectively change the macromodel of the lens systems -- e.g., shifting the lens center or changing the overall ellipticity of the lens.  The light gray region shows where the final 1\% of $R_{\rm cusp}$ values lie for the substructure-included systems.  Here, some large deviations from the no substructure case can be seen.  In particular, some small number of lenses show $R_{\rm cusp} \sim 1$ at opening angles less than $90\degr$.  So large deviations from the cusp relation due to the influence of substructure are possible but are infrequent.
 
We bin the cusp relation values by opening angle and focus on three bins that contain observed cusp systems:  
$30\degr \le \Delta \theta \le 40\degr$, $70\degr \le \Delta \theta \le 80\degr$, and $100\degr \le \Delta \theta \le 110\degr$.  
When comparing our mock lens systems to observed $R_{\rm cusp}$ values, we find that the probability of high $R_{\rm cusp}$ values remains small.  A summary is presented in Table \ref{tab:results}.  Here we use all 200 substructure realizations and estimate the cumulative probability of observing $R_{\rm cusp}$ values that are greater than provided by smooth models, $P(R_{\rm cusp} > RS)$, where $RS$ is the $R_{\rm cusp}$ value in the no substructure case where $P(R_{\rm cusp} > RS | {\rm no ~subhalos})=0$.  For mock lens systems using the substructure realizations, $P(R_{\rm cusp} > RS) \approx10^{-2}$.  In addition, the cumulative probability of observing $R_{\rm cusp}$ values that are greater than the observed values, $P(R_{\rm cusp} > RO)$ is estimated, where $RO$ is the largest $R_{\rm cusp}$ value observed for a given bin of opening angle.  For mock lens systems using the substructure realizations, $P(R_{\rm cusp} > RO) \approx10^{-3}$. 

In Fig. \ref{fig:conc}, we show the distribution of $R_{\rm cusp}$ values in a given bin of opening angle.  Here, the distribution of values for mock lenses with no substructure cluster near small $R_{\rm cusp}$ values  (shown as cyan bars).  On the other hand, the observed values are found at large $R_{\rm cusp}$ values (cross-hatched, magenta bars).  The distribution of values for mock lenses with substructure are shown in red bars and black-lined bars.  Substructure induces high-$R_{\rm cusp}$ tails which extend to (and, in some cases, past) the observed values of the cusp relation.  In all three opening angle bins, these tails contain only a small fraction of the lens systems (a few percent).  Interestingly, the high-$R_{\rm cusp}$ tails extend to larger values of $R_{\rm cusp}$ for smaller opening angles than for larger opening angles.  The substructure realizations are split by the concentration of the host galaxy, as discussed in Sec.~\ref{sec:sa}, plotting only the realizations that constitute the smallest 16\% of host concentrations (red bars) and the realizations that make up the largest 16\% (black-lined bars).  We see that the correlation between host concentration and amount of substructure persists into observable lensing results.  The hosts with the smallest concentrations have the largest amount of substructure and larger values of $R_{\rm cusp}$ when compared to the hosts with the largest concentrations.  This is best seen in the bins of $70-80\degr$ and $100-110\degr$, where the number of mock lens systems  number in the thousands.  In the smallest opening angle bin of $30-40\degr$, the number of mock lens systems is less than 20 per substructure realization and differences based on halo concentration are more difficult to discern.

We test which parameterizations of the substructure content of lensing galaxies are most sensitive to anomalous fluxes and are most useful for comparing observed lens systems to the results of simulations.  First, we explicitly tailor the definition of $f_{\rm sub}$ from Section \ref{sec:sa}, Eq. \ref{eqn:fsub} to match our macromodel conditions:  
\begin{equation}
f_{\rm sub} (<2b) = \frac{1}{M_{\mathrm{lens}} (R< 2b)} \sum_i M_{i} (R_{i}< 2b) ,
\label{eqn:fsub_2b}
\end{equation}
where $M_{i} (R_{i}< 2b)$ is a subhalo with mass $M$ that is found to be projected within twice the Einstein radius of the lens $b$ (within which all the images using macromodels M1, M2, and M3 have been found) and $M_{\mathrm{lens}}(R<2b)$ is the mass of the lensing galaxy within $2b$.  In this equation, we have the freedom to define what it means for a subhalo to be projected within $R=2b$.  As discussed previously, the substructure in our catalogs have density profiles which are described by a truncated NFW profile, which has three parameters:  the concentration, the bound mass, and the tidal radius.  We could chose to include any subhalo which is projected such that some portion of the area within $R=2b$ falls within the {\it tidal} radius of the subhalo.  This parameterization is supported by \citet{rozo_etal06}, who suggest that the most important DM halos for magnification perturbations are those within a tidal radius of an image.  Another possibility is that a single subhalo could be projected such that some portion of the area within $R=2b$ falls within the {\it scale} radius of the subhalo.  A final possibility is that a single subhalo could be projected such that the center of the subhalo falls within the area defined by $R=2b$.  This is the parameterization used in previous studies, when analyzing the outputs of dark matter simulations, \citep{mao_etal04,xu_etal09}.  The list of different parameterizations, then, is as follow:   1.)  include subhalos whose tidal radii fall within $2b$;  2.) include subhalos whose scale radii fall within that area; and 3.) include subhalos whose halo centers fall within that area.  

In Fig. \ref{fig:cuts}, the substructure realizations are split into low $f_{\rm sub}$ and high $f_{\rm sub}$ bins using the three criteria.  Note that some realizations have {\it no} subhalos which satisfy the criterion, i.e., $f_{\rm sub}=0$:  4 out of 200 for case 1, 45 out of 200 for case 2, and 58 out of 200 for case 3.  In addition, note that values of $\fsub \geq 1$ are possible.  This may be the result of two effects:   $M_{\mathrm{lens}} (R< 2b)$ from Eq. \ref{eqn:fsub_2b} only refers to the smooth mass of the host halo and not the total mass, and we use the full subhalo mass even when most of the subhalo's bound mass falls outside of the relevant region.  Using all three criteria, we can see that substructure realizations with high $\fsub$ values also have larger and more frequent violations of the cusp relation, when compared to low $\fsub$ realizations.   The differences between low and high values of $\fsub$ are greatest for values of $f_{\rm sub}$ calculated via halo centers (case 3).  Here, the median $\fsub$ value is $f_{\rm sub} =0.13\%$, and substructure realizations with $f_{\rm sub}$ greater than the median have an appreciable but small ($\sim 10^{-2}$) fraction of large $R_{\rm cusp}$ values while those with $f_{\rm sub}$ smaller than the median may have no large cusp relation values.  

The value of the cusp relation for every mock lens system is dependent on the relative positions of the subhalos and the source.  The placement of the subhalos and the source is stochastic -- depending upon the merger history of the lens halo, the direction in which the halo is projected, and the sampling of the source plane.  Thus, the results of Fig. \ref{fig:cuts} could be due to just a few substructure realizations that always give high-$R_{\rm cusp}$ values, while the majority of substructure realization give small $R_{\rm cusp}$ values regardless of $\fsub$.  We address this possibility in  Fig. \ref{fig:T345}.  For every bin of $f_{\rm sub}$, we measure the fraction of substructure realizations that have at least one large cusp relation value among their set of mock lenses.   A small fraction of the realizations with small values of $\fsub$ have large $R_{\rm cusp}$ values, while a large fraction of the realizations with large values of $\fsub$ have large $R_{\rm cusp}$ values.  This behavior is most pronounced for a $f_{\rm sub}$ definition that includes only subhalos whose center falls near the image positions, even when considering that, using this definition, the bins with $\fsub > 0.05$ contain only a few realizations each.  Thus, $\fsub$ is reliably correlated with $R_{\rm cusp}$ values. 

The parameterization of substructure in lensing halos which is most often used in studies of anomalous flux ratios is $\fsub$.  Other possibilities for parameterizing substructure exist:  the number of subhalos and the average subhalo mass.\footnote{Note that the average subhalo mass is not uniquely defined because it scales with the limits of the subhalo mass function, i.e., the average mass can be changed by changing the lower limit of subhalo masses investigated.  The lower limit in our work is $10^{-5} M_{\rm lens}$. }  We test these possibilities in Fig. \ref{fig:cuts2}.  Here,  subhalos whose halo centers fall within $R=2b$ are counted and the mean subhalo mass measured.  The substructure realizations with a larger number of subhalos also have larger and more frequent violations of the cusp relation, when compared to realizations with a smaller number of subhalos.  In addition, the substructure realizations with a larger average subhalo mass have larger and more frequent violations of the cusp relation, when compared to realizations with a smaller average subhalo mass.  However, when compared to Fig. \ref{fig:cuts}, neither the number of subhalos or the mean subhalo mass is as discriminating as $f_{\rm sub}$.  For example, for the observed value of  $R_{\rm cusp}=0.25$ (for opening angle bin $70-80\degr$), Figure \ref{fig:cuts2} suggests that the mean subhalo mass and the number of subhalos could be large or small.  On the other hand, 
Figure \ref{fig:cuts} is clear in showing that $\fsub$ must be large.  This result is consistent with that of \citet{metcalf_amara10}, who find that the frequency of flux anomalies is a function of $f_{\rm sub}$ and not of the number or average mass separately.  

\section{Caveats}
\label{sec:caveats}

In this section we discuss some specific caveats to our calculation that we are aware of and should be taken into account in the interpretation of these results. 


In Sec.~\ref{sec:strong}, we discussed how the simulations of substructure differ between the semi-analytic host halo and the typical lensing halo and the accommodations we make in order to incorporate the substructure catalogs into our mock lenses.  We make simple adjustments -- adopting the projected distance distribution and projected mass fraction from the simulations while scaling the mass of the host galaxy from $M_{\rm h} = 1.8 \times 10^{12} M_{\sun}$ to $M_{\rm lens} = 10^{13} M_{\sun}$.  The scaleability of the halo and subhalo masses is in line with the results of dark matter numerical simulations \citep{gao_etal04,giocoli_etal10}.  The host halo in the semi-analytic realizations also differs with respect to typical lensing halos in more subtle ways: the formation and merger history and the epoch of observation \citep{madau_etal08}.  While these differences are not likely to significantly effect the distance distribution or mass fraction of substructure \citep[e.g.,][and as discussed in Sec. \ref{sec:strong}]{xu_etal09}, they have unquantified effects on the internal structure of the subhalos -- their scale and tidal radii.  The concentration of a subhalo, however, is a slowly varying function of mass and time, so that 
biases in this analysis induced by these effects are likely to be small compared to natural 
halo-to-halo variation.


With regards to finite source effects, in this work we use the point source approximation, which -- as stated previously -- could bias the calculated magnification of high magnification images.  For example, \citet{metcalf_amara10} suggest that finite source effects could have large effects on the magnifications of images.  For lens systems with substructure and total magnifications above $\mu = 10$, the fractional deviation in magnification for a point source and for a 1 pc source approaches 1.  For smooth lens halos, the same holds true for magnifications above $\mu=20$.  The bias in magnifications most often manifests itself as an overestimate of the magnification of an image when a point-like source is used in place of a finite source.  

However, tests using our code fail to find significant biases due to the point-like source assumption.  This difference can be accounted for by two effects.  First of all, we consider only the cusp relation, which is the difference between magnifications of images, mitigating the effects of any finite source effect.  Secondly, \citet{metcalf_amara10} use a model of substructure where the surface density profile is a power-law $\Sigma \propto R^{-\alpha}$, and $\alpha=1$ or $\alpha=0.5$.  This profile is significantly steeper than the NFW profile used in this work and is likely to cause bigger magnification perturbations than the subhalos in our work.  


Finally, with regards to magnification bias, the distribution of opening angle values in our mock lenses, as shown in Figs. \ref{fig:rcusp_nosubs} and \ref{fig:rcusp_m}, is not representative of the distribution expected in observed lens samples.  For example, our mock data set has fewer lenses with small opening angles than expected observationally.  This is because the distribution of opening angles in our mock samples is a result of a combination of the macromodel parameters and the manner in which the source plane is sampled in order to create the mock lens systems.  We sample the source plane uniformly, a method which does not account for the magnification bias of observations -- the fact that brighter systems are more likely to be observed.  Following \citet{keeton_zabludoff04}, it can be shown that the probability density in the source plane is related to the luminosity function of sources, such that
\begin{equation}
p({\bf u})d{\bf u} = \mu({\bf u})^{\nu-1} d{\bf u},
\end{equation}
when the luminosity function of sources has a power-law form $dN/dS \propto S^{-\nu}$, where $S$ is the source flux and where $\bf{u}$ is the position of the source.  Thus if $\nu=2$, the probability of observing a lens system with source position $\bf{u}$ is proportional to the total magnification of the system, $\mu$.  Studies of the Cosmic Lens All-Sky Survey \citep[CLASS;][]{browne_etal03} have suggested that $\nu=2.1$ \citep{rusin_tegmark01,chae03}.  The total magnification of a system  is related to the opening angle of a cusp so that the smallest opening angles are found in the systems with the largest magnifications.  In our observational sample, one of ten in the observational set has an opening angle less than $40\degr$.  The mock lens systems, in contrast, have very few systems ($< 1\%$) with opening angles less than $40\degr$.  Accounting for magnification bias would increase the number of systems with opening angles less than $40\degr$ to a few percent.  Given the effect of magnification bias, we negate its effect by making comparisons of observed lens systems to our mock lens systems using only data that is binned in opening angles.

\section{Conclusions and Discussion}
\label{sec:conclusions}

Dark matter substructure is a generic prediction of the cold dark matter model and probes of substructure are a key test of the 
nature of the dark matter.  Strong gravitational lensing can be used to detect and measure dark matter subhalos, as they cause perturbations to the image magnifications, positions, and time delays in lens systems with galaxy lenses.  Putting robust constraints on these observational results require robust predictions for CDM substructure that account for halo-to-halo variation.  Current predictions from numerical simulations have used a limited number of simulated halos.  We investigate the question of halo-to-halo variation in substructure using semi-analytical models and test how it affects perturbations to the magnifications of images, as measured by the cusp relation parameter, $R_{\rm cusp}$.  We find the following results.
\begin{enumerate}

\item The total projected substructure mass fraction for halos with mass greater than $10^{-5}$ of the mass of the host is $\sim 9\%$ with a 95 percentile in the range $[2-17]\%$.  Low concentration halos have a higher projected mass fraction ($\sim 20\%$), with a significantly larger 95 percentile range. High concentration halos have a lower mass fraction ($\sim 5\%$), and a much smaller 95 percentile range. 

\item We find that the mass fraction within the inner 3\% of the projected radius is $f_{\rm sub}=0.25\%$, with a 95 percentile in the range of $f_{\rm sub}=[0-1]\%$ for the complete sample of 200 realizations. For low-concentration host halos the median substructure mass fraction is 
$f_{\rm sub} \sim 0.5\%$ with a 68 percentile in the range of $f_{\rm sub}=0.2-0.8\%$.  These results are generally consistent with the 
results of \citet{xu_etal09}, who find $.01\% \leq f_{\rm sub} \leq .7\%$, using 6 halos from the Aquarius simulation,  as well as with all previously published limits on $\fsub$.  The mass fraction of host halos with higher concentrations is lower, with $f_{\rm sub} \le 0.1\%$ for host halos 
with concentrations in the upper 16\% of the distribution.  

\item Accounting for halo-to-halo variation, the observed values of $R_{\rm cusp}$ are greater than the values predicted by smooth lens models, and larger 
than 99\% of the values predicted by substructure models.  Given halo-to-halo variation, the probability of observing cusp relation values greater than supported by smooth models is $\sim$$10^{-2}$ and the probability of observing cusp relation values greater than the largest currently observed values is $\sim$$10^{-3}$.

\item Substructure fractions are correlated with the concentrations of the host halos.  Lower-concentration host 
halos have larger substructure fractions so that lower concentration host halos give rise to more substructure 
and more violations of the cusp relation.

\item  For lensing studies of substructure, the substructure content of lensing halos is best parameterized by a definition of the substructure mass fraction, $\fsub$, which includes only the subhalos whose halo centers fall within the area in which images are expected to be found (twice the Einstein radius in our models, $\sim3\%$ of the virial radius of the lens halo, generically) and which excludes more distant subhalos that fall within a scale or tidal radius of a lensed image.

\item  $f_{\rm sub}$ is a better predictor of $R_{\rm cusp}$ when compared to the number of subhalos or the mean subhalo mass.

\end{enumerate}

We attempt to isolate the effect of halo-to-halo variation in substructure populations from macromodel effects and, thus, do not model any particular observed lens system.  Our efforts can still be compared qualitatively to previous attempts to constrain the substructure mass fraction.  For example, \citet{dalal_kochanek02} find that $f_{\rm sub}=0.02$ with $.6\% < f_{\rm sub} < 7\%$ within 90\% confidence intervals.  This result is marginally consistent with predictions of $f_{\rm sub}$ in previous works and in this paper, while \citet{metcalf_amara10} find rough consistency with such predictions.  Both of these works use substructure models with density profiles that are steeper than the NFW profile used in this work.  Steeper profiles are more effective at causing magnification perturbations.  Thus, it is possible that both these works are underestimating the $f_{\rm sub}$ that is supported by lensing data.  Other works have also suggested that lensing observations are inconsistent with expectations from numerical simulations  \citep{maccio_etal06,amara_etal06,chen09}.  \citet{xu_etal09} suggest that, using the 6 halos in the Aquarius simulation, the probability of the observed values of $R_{\rm cusp}$ is $\sim$$10^{-3}$.  

In summary, we find that when accounting for halo-to-halo variation the probability of observing $R_{\rm cusp}$ values that match observations may be as large as $\sim$$10^{-2}$, but the tension between observations and numerical simulations persist.  This tension may be eased by assuming larger values of the substructure mass fraction.  Possible solutions, then, are 1.) additional substructures which are unresolved in current numerical and semi-analytic simulations and 2.) biases which cause the observational sample to prefer lensing halos with high substructure mass fraction or, relatedly, low concentration.  Both these possibilities, however, are speculative.  

While this work only probes the effect of subhalos with masses $\sim$$10^{8} M_{\sun}$ or larger, others have studied smaller substructures.  Using a 
$\Lambda$CDM model, \citet{maccio_miranda06} investigated the role of subhalos with masses $10^{5} - 10^{7} M_{\sun}$ and found them to be insufficient in creating large cusp relation violations.  The Aquarius simulation used by \citet{xu_etal09} resolves subhalos down to $10^{5} h^{-1} M_{\sun}$. The cumulative effect of smaller and larger subhalos produces their result of a probability that observations match expectations from $\Lambda$CDM of $\approx 10^{-3}$.  

Currently, there are no known observational biases which would prefer high substructure or low concentration lensing halos.  If the observational bias from the concentration of lensing halos is considered alone, high concentration halos should be preferred \citep{mandelbaum_etal09}.  In addition, the relation between observational biases, substructure content, and other characteristics of lenses may be complicated.  For example, triaxial halos projected along the long axis are better strong lenses in general and may have more substructure projected near images \citep{zentner06}, but triaxial halos along the middle axis are better at creating four-image lenses \citep{rozo_etal07}.  Accounting for all the biases due to lensing halo characteristics is difficult and requires future investigation.

\begin{acknowledgements}
SMK is funded by NSF grant PHY-0969853 and by Brown University. 
ARZ is funded by the National Science Foundation under grant PHY 0968888.  
\end{acknowledgements}

\bibliography{manuscript}

\end{document}